\documentclass[paper]{JHEP3}
\usepackage{epsfig}
\usepackage{graphicx}
\usepackage{xspace}
\usepackage{amssymb}
\usepackage{subfigure}
\usepackage{multirow}

\newcommand{\be}{\begin{equation}}
\newcommand{\ee}{\end{equation}}

\newcommand{\bea}{\begin{eqnarray}}
\newcommand{\eea}{\end{eqnarray}}
\newcommand{\beanon}{\begin{eqnarray*}}
\newcommand{\eeanon}{\end{eqnarray*}}
\newcommand{\ba}{\begin{array}}
\newcommand{\ea}{\end{array}}
\newcommand{\bd}{\begin{description}}
\newcommand{\ed}{\end{description}}
\newcommand{\bi}{\begin{itemize}}
\newcommand{\ei}{\end{itemize}}
\newcommand{\ben}{\begin{enumerate}}
\newcommand{\een}{\end{enumerate}}
\newcommand{\bc}{\begin{center}}
\newcommand{\ec}{\end{center}}

\newcommand{\ifb}{\mbox{ fb}^{-1}}

\newcommand{\pt}{\mbox{${\mathrm p_T}$}\xspace}

\newcommand{\ordEW}{\mathcal{O}(\alpha_{\scriptscriptstyle EM}^6)\xspace}
\newcommand{\ordQCD}{\mathcal{O}(\alpha_{\scriptscriptstyle EM}^4
  \alpha_{\scriptscriptstyle S}^2)\xspace}
\newcommand{\ordQCDsq}{\mathcal{O}(\alpha_{\scriptscriptstyle EM}^2
  \alpha_{\scriptscriptstyle S}^4)\xspace}
\newcommand{\ordQCDtwo}{\mathcal{O}(\alpha_{\scriptscriptstyle EM}^2
  \alpha_{\scriptscriptstyle S}^4)}

\newcommand{\eqn}[1]{Eq.(\ref{#1})}

\newcommand{\eqnsc}[2]{Eqs.(\ref{#1},~\ref{#2})}
\newcommand{\tbn}[1]{Tab.~\ref{#1}}

\newcommand{\fig}[1]{Fig.~\ref{#1}}

\newcommand{\sect}[1]{Sect.~\ref{#1}}

\newcommand{\rf}[1]{Ref.~\cite{#1}}
\newcommand{\rfs}[1]{Refs.~\cite{#1}}  % comma separated list

\newcommand{\Phantom}{{\tt PHANTOM}\xspace}

\newcommand{\MadEvent}{{\tt MADEVENT}\xspace}

\def\pl #1 #2 #3 {{\it Phys.~Lett.} {\bf#1} (#2) #3}   
\def\np #1 #2 #3 {{\it Nucl.~Phys.} {\bf#1} (#2) #3}
\def\zp #1 #2 #3 {{\it Z.~Phys.} {\bf#1} (#2) #3}
\def\pr #1 #2 #3 {{\it Phys.~Rev.} {\bf#1} (#2) #3}
\def\prep #1 #2 #3 {{\it Phys.~Rep.} {\bf#1} (#2) #3}
\def\prl #1 #2 #3 {{\it Phys.~Rev.~Lett.} {\bf#1} (#2) #3}
\def\intj #1 #2 #3 {{\it Int. J. Mod. Phys.} {\bf#1} (#2) #3}
\def\mpl #1 #2 #3 {{\it Mod.~Phys.~Lett.} {\bf#1} (#2) #3}
\def\rmp #1 #2 #3 {{\it Rev. Mod. Phys.} {\bf#1} (#2) #3}
\def\cpc #1 #2 #3 {{\it Comp. Phys. Commun.} {\bf#1} (#2) #3}
\def\epj #1 #2 #3 {{\it Eur. Phys. J.} {\bf#1} (#2) #3}
\def\jhep #1 #2 #3 {{\it JHEP} {\bf#1} (#2) #3}

\title{
Vector-Vector scattering at the LHC with two charged leptons and two neutrinos
in the final state.
}

\author{
Alessandro Ballestrero$^a$,
Diogo Buarque Franzosi$^{a,b}$ and
Ezio Maina$^{a,b}$\\
$^a$ INFN, Sezione di Torino, Italy,\\
Via Giuria 1, 10125 Torino, Italy,\\
$^b$ Dipartimento di Fisica Teorica, Universit\`a di Torino, Italy\\
Via Giuria 1, 10125 Torino, Italy.\\
Email: ballestrero@to.infn.it, buarque@to.infn.it, 
maina@to.infn.it.
}

\preprint{DFTT 15/2010}

\abstract{
A complete parton level analysis of $2\ell 2\nu 2j$ and $4\ell 2j$, $\ell = \mu,\, e$
production at the LHC is presented, including all processes at order
$\ordEW$, $\ordQCD$. The infinite Higgs mass
scenario, which is considered as a benchmark for strong scattering theories and
is the limiting case for composite Higgs models, 
and one example of 
Strongly Interacting Light Higgs models 
are confronted with the
Standard Model light Higgs predictions. This analysis is combined with the
results in the $\ell\nu$ + four jets, the $\ell^-\ell^+$ + four jets 
and the $3\ell\nu$ + two jets channels
presented in  previous papers, in
order to determine whether
these alternative Higgs frameworks
can be detected as an excess
of events in boson--boson scattering.
}

\begin{document}

\section{Introduction}
\label{sec:intro}

Whether or not the search for a light Higgs boson at the LHC will be successful,
vector boson scattering processes will require careful analysis.
In fact, the corresponding amplitudes involving only vector bosons 
grow with energy when the bosons are longitudinally polarized
and violate perturbative unitarity at
about one TeV, requiring either the Higgs or some new physics 
in the energy range accessible to the LHC in order to tame this unphysical
behaviour\footnote{ Detailed reviews and extensive bibliographies can be found in
Refs.~\cite{HiggsLHC,djouadi-rev1,ATLAS-TDR,Houches2003,CMS-TDR,Chanowitz:1998wi}}.

The Standard Model (SM) describes Electroweak Symmetry Breaking (EWSB)
through a single complex Higgs doublet. 
Many alternative mechanisms of EWSB however
have been explored. We will not try to
summarize the different models and simply refer to the literature.
We will only remark that it is conceivable
that composite states are responsible for EWSB
\cite{H_Goldstone,LittleH1,LittlestH,gaugeHiggsU1,gaugeHiggsU2,HologHiggs,LittleHcustodial,Giudice:2007fh}
.
These theories typically predict the presence
of new states which, if light enough, could be observed at the LHC.

The effective field theory approach \cite{EEWL,Contino07,Giudice:2007fh,Barbieri}
is a powerful method for treating the
low energy dynamics of systems with broken symmetries. It provides a systematic
expansion of the full unknown Lagrangian in terms of the fields which are
relevant at scales much lower than the symmetry breaking scale. 

In Ref.~\cite{Giudice:2007fh}
it has been pointed out that, if EWSB is
triggered by a light composite Higgs which is a pseudo--Goldstone boson related
to some large scale strongly interacting dynamics, the growth with energy of the
vector boson scattering amplitudes typical of Higgsless models might not be
completely canceled by Higgs exchange diagrams but only slowed down.
This kind of models have been called Strongly Interacting Light Higgs (SILH) models.
Examples which fall into this class are for instance the
Holographic Higgs \cite{HologHiggs}, the
Little Higgs of Ref.~\cite{LittleHcustodial} and the
Littlest Higgs \cite{LittlestH}.

In SILH models the leading low energy effects are described by
two parameters (one responsible for a
universal modification of all Higgs couplings, and the other one for a universal
modification of Higgs couplings to fermions) characterized by the ratio $v^2/f^2
= \xi$, where $v$ is the Higgs vacuum expectation value and $f$ is the
$\sigma$--model scale.
The natural range of the $\xi$ parameter is between $\xi=0$ and $\xi=1$ which correspond
respectively to the limiting cases of the Standard Model and of technicolor theories.
Because of the modified Higgs couplings, longitudinal
gauge--boson scattering amplitudes violate unitarity at high energy, even in the
presence of a light Higgs \cite{Giudice:2007fh}.

%%%%%%%%%%%%%%%%%%%%%%%%%%%%%%%%%%%%%%%%%%%%%%%%%%%%%%%%%%%%%%%%%%%%%%%%%%%%%%%%%
%the large number of different proposals it is useful to determine the model
%independent features of this class of theories.
%There has been recent progress in this area \cite{Contino07,Giudice:2007fh,Barbieri},
%using the effective
%theory language \cite{EEWL}. 
%If the mass of the Higgs boson becomes larger than the typical energy
%scale at which boson--boson scattering is probed, the contribution of the Higgs
%exchange diagrams decreases and completely vanishes, in the Unitary gauge we
%employ throughout our calculation, in the limit of an infinite mass which we
%will refer to as the Higgsless case. As a consequence, the scattering cross
%section for an infinitely massive Higgs in the SM represents, at large energies,
%an upper limit for VV scattering processes in SILH models, and can be taken as a
%benchmark for the observability of signals of strong scattering and of Higgs
%compositeness in boson--boson reactions. On the other hand the Higgsless case is
%also representative of models in which heavy resonances which
%unitarize boson--boson scattering  are present but cannot be directly detected at the
%LHC.

Scattering processes among vector bosons have been scrutinized since a long time
\cite{history1}. In Ref.~\cite{Accomando:2005hz,Accomando:2006vj}
an analysis of $\ell\nu$ + four jets and $\ell^+\ell^-$
+ four jets production at the LHC has been presented, with the limitation of
taking into account only purely electroweak processes. Preliminary results
concerning the inclusion of the $\ordQCD$ background, which include $V V + 2j$ and
top--antitop production have appeared in Ref.~\cite{Ambroglini:2009mg}.
A preliminary analysis in the Equivalent Vector Boson Approximation of the observability
of partial unitarization of longitudinal vector boson scattering in SILH models
at the LHC can be found in Ref.~\cite{Cheung:2008zh}.
In the last few years QCD
corrections to boson--boson production via vector boson fusion \cite{JagerOleariZeppenfeld}
at the LHC
have been computed and turn out to be below 10\%. Recently, VBFNLO \cite{Arnold:2008rz}
a Monte
Carlo program for vector boson fusion, double and triple vector boson production
at NLO QCD accuracy, limited to the leptonic decays of vector bosons, has been
released.
Recently, the first results for the NLO corrections to $W+4j$ production
have appeared \cite{Berger:2010zx}.
New techniques which exploit the angular distribution of vector boson decay products
to determine the ratio of longitudinal and transverse polarization have been proposed
in \cite{Han:2009em}.

In \rf{Ballestrero:2008gf}
a complete parton level analysis of $\ell\nu$ + four jets production
at the LHC, including all processes at order $\ordEW$, $\ordQCD$ and $\ordQCDsq$
has been presented, comparing a typical SM light Higgs scenario with the
Higgsless case. It was noted that the $\ordQCDsq$ $W$ + 4j background is so
large that the usual approach of comparing the number of events in the two
scenarios at large invariant masses is useless.
It was argued that the invariant mass distribution of the two
central jets in the vector--vector scattering signal
presents a peak corresponding to the decays of
vector bosons while
the background produced by $\ordQCDsq$ $W +4j$ processes is rather flat and
therefore can be measured from the sidebands and subtracted,
drastically decreasing the theoretical uncertainties.

In \rf{Ballestrero:2009vw} the processes $pp
\rightarrow \ell^+\ell^- + 4j$ and $pp \rightarrow 3\ell\nu + 2j$
have been studied along the lines introduced in \rf{Ballestrero:2008gf}.
The infinite mass Higgs scenario and the instance of SILH models described above
have been compared with the light Higgs SM framework.

In this paper we concentrate on the boson boson scattering reactions which
produce a $2\ell 2\nu 2j$ final state. Because of the presence
of two neutrinos the mass of the final state boson pair cannot be reconstructed.
For completeness sake in the end we also discuss the $2j4\ell$ channel in which
the vector pair mass can obviously be measured with high accuracy but which has
been left out of our previous papers because of its small cross section.
These processes have been studied already in \rf{history2}
where they have been described as gold--plated.
A potentially large background to these channels
is the copious yield of high \pt, isolated leptons in B--hadron production
\cite{iso-lept-from-B} which mimic the signature of the leptonic decays
of $W$ bosons. A detailed experimental analysis of two same sign $W$'s
has however shown that when standard isolation criteria are applied isolated
leptons from B--hadrons can be efficiently eliminated \cite{Zhu:2010cz}.
Therefore, we have reanalyzed the $2\ell 2\nu 2j$ channels using complete
$\ordEW$ and $\ordQCD$ samples.

We have 
estimated the probability that, assuming that either the
Higgsless scenario or the instance of SILH model we have considered is realized
in Nature, the results of the measurements 
at the LHC yield results which are incompatible with the SM.
We have first
combined separately the three channels,
$2j\ell^\pm\ell^\pm\nu\nu$, $2jZZ\rightarrow 2j\ell\nu\ell\nu$,
$2jWW\rightarrow 2j\ell\nu\ell\nu$ in which the invariant mass of the final state
cannot be directly measured and the four channel,
$2j4\ell$, $4j\ell\nu$,$4j\ell\ell$ and $2j3\ell\nu$ in which it can instead be
reconstructed. Finally we have combined all channels.

%%%%%%%%%%%%%%%%%%%%%%%%%%%%%%%%%%%%%%%%%%%%%%%%%%%%%%%%%%%%%%%%%%%%%%%%%
\section{Calculation}
\label{sec:calculation}

Two perturbative orders contribute to the $2\ell 2\nu 2j$ and $4\ell 2j$ signals
at the LHC.
The purely electroweak set of diagrams at $\ordEW$ is the one which includes
boson boson scattering as a subprocess. In the second set of diagrams at $\ordQCD$
no such scattering takes place: either two fermion lines exchange a gluon or
a single fermion line and two external gluons are present.  
In contrast with the processes examined in \rfs{Ballestrero:2008gf,Ballestrero:2009vw}
where the dominating background is due to $V+4j$ $\ordQCDtwo$ processes
in which only one vector boson is produced,
in the present case the final states from both perturbative orders contain 
two vector bosons and are essentially impossible to separate.
In \rfs{Eboli:2006wa,Englert:2008tn} it has been pointed out that $t\overline{t}$+$n$--jets
production, $n=1,2$ can
provide a significant background to vector boson scattering.
Indeed, the additional jets which are present in the former processes can go undetected
and mimic the signature of boson boson reactions.

Both the $\ordEW$ and the $\ordQCD$
samples have been generated with \Phantom , a dedicated
tree level Monte Carlo generator
which is documented in Ref.~\cite{Ballestrero:2007xq} while additional material
can be found in Refs.~\cite{ref:Phase,method,phact}.
The $t\overline{t}$+$n$--jets processes have been simulated with \MadEvent
\cite{MadeventPaper} in the Narrow Width Approximation.
Both programs generate events in the Les Houches Accord File Format \cite{LHAFF}.
For the LHC we have assumed the design energy of 14 TeV.
For each perturbative order we have generated a sample of five hundred thousand
unweighted events. In some cases additional event samples have been generated
in order to increase the final statistics of particular phase space regions.

%For the Standard Model parameters we use the input values:
%\begin{equation}
%\begin{array}[b]{lcllcllcl}
%%\begin{eqnarray}
%\label{eq:SMpar}
%\mathrm {M_W} & = & 80.40 , \qquad &
%\mathrm {M_Z} & = & 91.187~\mathrm{GeV}, \\
%G_{\mu} & = & 1.16639~10^{-5}~\mathrm{GeV}^{-2}, \qquad &
%\alpha_s(\mathrm {M_Z})  & = & 0.118\\
%\mathrm {M_t} & = & 175.0~\mathrm{GeV}, \qquad &
%\mathrm {M_b} & = & 4.8~\mathrm{GeV}. 
%%\begin{eqnarray}
%\end{array}
%\end{equation}
%
%The masses of all other partons have been set to zero. We adopt the standard
%$G_{\mu}$--scheme to compute the remaining parameters.

All samples have been generated using CTEQ5L \cite{CTEQ5} parton distribution
functions. The QCD scale has been taken as
\be
\label{eq:LargeScale}
\Lambda^2 = M_W^2 + \frac{1}{6}\,\sum_{i=1}^6 p_{Ti}^2
\label{scale}
\ee
in all cases but for the reaction in which
a triplet of final state particles with flavours compatible with deriving
from the decay of a top or antitop quark could be found.
In this case the scale has been taken as
\be
\label{eq:LargeScaleTop}
\Lambda^2 = M_{top}^2 + p_{Ttop}^2.
\ee

The analysis has been performed
at parton level with no showering and hadronization.

The effective Lagrangian approach to SILH models of Ref.~\cite{Giudice:2007fh} 
is valid for small
values of $\xi$,  while larger values demand a more detailed description of the
particular model at hand. Such a Lagrangian leads to a modification of the Higgs 
couplings by a factor $1/\sqrt{1+c_H\xi}$, which can be reabsorbed in a Higgs
propagator modification by a factor $1/(1+c_H\xi)$ in boson boson scattering
studies. $c_H$ is a pure number of order unity 
\cite{LittlestH,HologHiggs, LittleHcustodial,Giudice:2007fh}.
For the present study we have selected the value $c_H\xi = 1$ which we intend
as a possible upper limit for the model independent lagrangian description of
Ref.~\cite{Giudice:2007fh}.
Both for the SM scenario and for the SILH model the Higgs mass has been fixed at
$ 200 \mbox{ GeV}$. Since we are interested in large invariant mass vector vector
scattering processes the actual value of the Higgs mass is irrelevant,
provided it is appreciably smaller than the invariant mass of the vector pair.

For very large Higgs masses, all Born diagrams with Higgs propagators become
completely negligible in the Unitary Gauge we work in. Therefore the no--Higgs
model results for all processes coincide with
those in the $M_H \rightarrow \infty$ limit.
This framework therefore can be considered
as an upper limit to SILH models and also as representative of all models in which
resonances which unitarize vector vector scattering
are present but too heavy to be directly detected.
The no--Higgs case is also extremely useful to determine the phase space regions
in which weak and strong interacting vector boson models differ the most.
It is obvious that the Standard Model with an infinite mass Higgs is unphysical
because of the violations of perturbative unitarity at about one TeV.
This corresponds in our complete calculation to an invariant mass of the two vector bosons
of the same magnitude.
These events are present at the LHC but they are highly suppressed by the
effective parton luminosities, as can be clearly extracted from the plots in 
\rfs{Ballestrero:2008gf,Ballestrero:2009vw} and in the following.
We have studied the behaviour of unitarized models, and in particular we have compared
the basic no--Higgs case with a model in which the no--Higgs amplitudes are
unitarized using the K--matrix method and with some models which contain resonances.
Typically the number of expected events in the presence of resonances is much larger
than for the no--Higgs framework.
What is perhaps more important, it turns out that, after cuts comparable to
the ones adopted in this paper, the expected number of events in the unitarized
no--Higgs model is only a few percent smaller than in the non--unitarized case.
Therefore we consider the possibility of distinguishing the no--Higgs case from the
SM at the LHC a quite solid benchmark for the observability of unitarized models.
The details of our studies of unitarized models are given in 
\rfs{BuarqueThesis,WorkInProgress}.

The selection of events with jets widely separated in pseudorapidity
is a well established technique for
enhancing the scattering contributions at the LHC \cite{history1}.
As shown in Ref.~\cite{Ballestrero:2008gf} a powerful tool to
increase the separation between the SM predictions and those of the Higgsless
scenario is provided, at large invariant masses,
by the request that the vector bosons and their decay
products are in the central part of the detector since the vector bosons in the
Higgsless case have smaller rapidities and larger momenta than in the presence
of a light Higgs.

The cuts in \tbn{tab:cuts_0} have been applied either at
generation level or as a preliminary step to any further analysis.
They require containment within the active region of the
detectors and minimum transverse momentum for all observed partons; a minimum
mass separation is imposed for all same--family opposite--sign charged leptons
and all jet pairs.
Furthermore, the two jets are required to be
separated by at least three ($2j2\ell 2\nu$) or four ($2j4\ell$)
units in rapidity and their combined mass is forced
to be outside the electroweak vector boson mass window in order to exclude
three vector boson production.

We have considered two different ranges for the mass of the lepton pair
in the $2j\ell^+\ell^-\nu\bar{\nu}$ channel. On one hand we have selected same flavour
charged leptons with a mass in the interval $76$ GeV $< M(\ell^+\ell^-) < 106$ GeV.
In this case we have considered the lepton pair to be produced in the decay
of a $Z$ boson. Requiring further a large missing transverse momentum we have 
produced an event sample corresponding to the
$2jZZ \rightarrow 2j\ell^+\ell^-\nu\bar{\nu}$ channel which will be discussed in 
\sect{sec:zz}.
When the mass of the lepton pair is outside the quoted range or the two
oppositely charged leptons belong to different families we consider the event a candidate
for the $2jWW$ channel. Since we are interested into high invariant mass $W$ pairs,
we have required $M(\ell^+\ell^-) > 250 \mbox{ GeV}$ for this kind of
events which will be analyzed in \sect{sec:ww}. The mass of the $WW$ system
corresponds to the scale of boson boson scattering and large masses help in discriminating
between the SM and other scenarios.

For both the Higgsless and SILH cases and for each channel
we have computed the probability that, assuming a specific Beyond Standard Model (BSM)
correctly describes nature, the result of an
experimental outcome for a given luminosity has a chance of less than 5\% in
the SM (PBSM@95\%CL).

For the combination of channels discussed in \sect{sec:combining}
we have also computed the 99.7\% exclusion limit (PBSM@99.7\%CL).

All limits presented in the following, unless explicitly mentioned, 
have been computed summing over all possible combinations
of first and second generation leptons, assuming an
integrated luminosity of $L=200 \ifb$, which 
we intend as corresponding to one year of high luminosity combining CMS and ATLAS
results.

%\begin{table}[ht!]
%\begin{center}
%\vspace{0.5cm}
\TABLE{
\begin{tabular}{|l|}
\multicolumn{1}{c}{ }\\
%\hline
\hline
$p_T(\ell^\pm) > 20 \mbox{ GeV}$ \\
\hline
$|\eta(\ell^\pm)| < 3.0$ \\
\hline
$M(\ell^+\ell^-) > 20 \mbox{ GeV}$\\
$M(\ell^+\ell^-) > 250 \mbox{ GeV}$ \quad ($2jW^+W^-$)\\
$76$ GeV $< M(\ell^+\ell^-) < 106$ GeV \quad ($2jZZ$)\\
\hline
$p_T(j) > 30 \mbox{ GeV}$ \\
\hline
$|\eta(j)| < 6.5$ \\
\hline
$M(jj) > 60 \mbox{ GeV}$ \\
$M(jj)<70 \mbox{ GeV} ; M(jj)>100 \mbox{ GeV}$ \\
\hline
$|\Delta \eta (jj)| > 3.0$ \quad ($2j2\ell 2\nu$)\\
$|\Delta \eta (jj)| > 4.0$ \quad ($2j4\ell$)\\
\hline
\end{tabular}
\caption{Acceptance cuts.
} 
\label{tab:cuts_0}
%\end{center}
%\end{table}
}

We proceed as follows.
We define the signal $S$ as the sum of the events for all $\ordEW$ and $\ordQCD$ processes
after all selection cuts. It might be feasible to further decrease the $\ordQCD$
contribution with a central jet veto, but this possibility is beyond the scope of this paper.
For the $2jWW\rightarrow 2j\ell\nu\ell\nu$ channel we take as background $B$ the
expected yield of the $t\overline{t}+jets$.
$B$ and $S$ are considered as random variables
representing the number of background and signal events for
a possible experimental outcome. $\overline{B}$ and $\overline{S}$ are the
corresponding average values which will be taken equal to the predictions of our
simulation.
We take into account the statistical uncertainty of
$S$ assuming a standard Poisson distribution with average
$\overline{S}$. The predicted signal cross
section is also affected by theoretical uncertainties, so the parameter
$\overline{S}$ is itself subject to fluctuations. 
The theoretical error is modeled by a flat
distribution in the window $\overline{S} \pm 30\%$ which, in our opinion, is a
reasonable choice to account for both pdf's and scale uncertainties for the
signal. The processes we are interested in require center of mass energies of
the order of the TeV and therefore involve quarks with rather large longitudinal
momentum fraction $x$, $x \approx
10^{-1} \div 10^{-2}$ at a typical scale $\Lambda$ of about 100 GeV. In this region
the uncertainty due to the parton distribution functions is of the order of
5\% \cite{Martin:2002aw,Martin:2003sk}. As already stated, QCD corrections are
in the range of 10\% and, as
a consequence theoretical uncertainties are expected to be well within this
order of magnitude.

Only statistical fluctuations have been taken into
consideration in the case of $B$. This is motivated by the fact that the $t\overline{t}+jets$
background is likely to be well measured experimentally in 
final states in which more than two jets are detected and then extrapolated
via Monte Carlo to the region of interest in this paper,
 so that the theoretical error on $B$ is not expected to be
an issue at the time when real data analysis will be performed.
In \sect{sec:combining} we will also discuss how our results would be affected if
$t\overline{t}+jets$ were not measured.

We define the
test statistics $D = S + B -\overline{B}$ which reduces to $S$ in the absence of background.
Having computed the probability distributions $P(D|SM)$ and $P(D|BSM)$ of $D$
in the Standard Model and in the Beyond the Standard Model under consideration,
the 95\%CL region for the SM can be defined from the probability ratio

\be
\label{eq:ratio}
Q(D)= \frac{P(D|BSM)}{P(D|SM)}
\ee

determining a number $\alpha$ such that

\be
\label{eq:alphaFix}
\int dD \, P(D|SM) \,\theta (\alpha - Q) = 95\%.
\ee

The
probability for the BSM to yield a result outside this 95\%CL region for the SM
is then

\be
\label{eq:PBSM}
PBSM@95\%CL = \int dD \, P(D|BSM) \,\theta (Q-\alpha). 
\ee

A number of comments, which apply to all channels discussed in this paper,
should be made.
We have performed a simple cut based study, which can undoubtedly be improved upon
with a more sophisticated multivariate analysis. On the other hand we have not
taken into account experimental efficiencies and all issues related to
additional hadronic activity due to showering and the underlying event.
The selection cuts discussed below have been chosen in order to maximize the
separation of the light Higgs case from the no--Higgs one.

In the following we will present cross sections as a function of appropriate
minimum invariant masses $M_{cut}$, typically extracted from lepton momenta.
The best discrimination between the SM and the BSM schemes are generally obtained for
$M_{cut}$ values which yield production rates which are uncomfortably small, 
particularly because of neglected experimental uncertainties.
It should however be noticed that at smaller values of $M_{cut}$ the rate is
usually much larger with a modest decrease of discriminating power.

%%%%%%%%%%%%%%%%%%%%%%%%%%%%%%%%%%%%%%%%%%%%%%%%%%%%%%%%%%%%%%%%%%%%%%%%%
\section{The $2j\ell^\pm\ell^{\prime\pm}\nu\nu$ channel: two same--sign leptons in the final state}
\label{sec:llss}

This channel, which is characterized by two same sign charged leptons, possibly
of different flavour, in the final state,
has low EW and QCD background, for no external gluons contribute to this 
final state.
We remark that the production of two same--sign $W$'s has been extensively discussed
in the context of Multiple Particle Interactions (MPI) 
\cite{Kulesza:1999zh,Maina:2009sj,Gaunt:2010pi}, since it has the peculiarity that
it can be realized in MPI at a lower perturbative order than in ordinary two
parton collisions where at least two additional partons must appear in the final state.
However, if two jets in the final state are required, the MPI contribution is small
and concentrated in the region of small total visible energy and therefore has been
neglected.

%\begin{table}[bt]
\TABLE{
\begin{tabular}{|p{0.08\textwidth}|p{0.27\textwidth}|p{0.27\textwidth}|p{0.26\textwidth}|}
\hline
$M_{cut}$ 	& \centering no Higgs 	& \centering SILH  & 
					\parbox[t]{0.26\textwidth}{\centering $M_H=200$ GeV}   \\ 
		\cline{2-4}
 (GeV) 		& \centering $\sigma$(fb) & \centering $\sigma$(fb) & \parbox[t]{0.26\textwidth}{\centering $\sigma$(fb)} \\
\hline
200             & 3.11(2.39) & 2.87(2.15) & 2.73(2.01)  \\
300             & 1.73(1.23) & 1.55(1.06) & 1.46(.967)  \\
400             & 1.01(.682) & .892(.560) & .839(.507)  \\
500             & .630(.407) & .538(.315) & .505(.283)  \\
600             & .400(.253) & .334(.187) & .311(.163) \\
700             & .262(.162) & .214(.114) & .198(.0975)  \\
800             & .177(.108) & .142(.0728) & .130(.0613) \\
\hline
\end{tabular}
\caption{Total cross section for the $\ell^\pm\ell^\pm\nu\nu+2j$ channel after generation cuts,
\tbn{tab:cuts_0}.
In parentheses the results for the $\ordEW$ sample. 
}
\label{tab:res_llss_0}
%\end{table}
}

The presence of two neutrinos in the final state
makes it impossible to reconstruct the invariant mass of
the di-boson system which corresponds to the center of mass energy 
of the $WW$ scattering.
For $M_{cut}$ we have resorted to a correlated observable, the di-lepton mass $M(\ell\ell)$.

The total cross section for the $2j\ell^\pm\ell^\pm\nu\nu$ channel with the acceptance cuts
in \tbn{tab:cuts_0} is presented in \tbn{tab:res_llss_0} as a function of the minimum 
$\ell\ell$ invariant mass $M_{cut}$. In parentheses the results for the $\ordEW$
processes. 
\tbn{tab:res_llss_0} shows that the cross section for the $\ordQCD$ processes
is only about  25\% to 40\% of the total cross section in the 
Higgsless scenario already at this level.
The distribution of the lepton pair mass,
with acceptance cuts only, is presented on the left hand side
of \fig{fig:llss_Mlltsum}. The $\ordQCD$ background is negligible at small di-lepton mass,
while it becomes of the same order of magnitude of the $\ordEW$ contribution for
$M_{cut} > 500\mbox{ GeV}$.

%\begin{table}[!ht]
%\centering
%\vspace{0.5cm}
\TABLE{
\begin{tabular}{|l|}
\hline
$\Delta \eta (jj) > 4.5$ \\
\hline
$max|\eta(j)|> 2.5$ \\
\hline
$|\eta(j)|> 1.$ \\
\hline
$|\eta(\ell)|<2.5$ \\
\hline
$p_T(\ell) >  50$ GeV \\
\hline
$min{p_T(j)} <  120$ GeV \\
\hline
$\Delta R (\ell j)> 1.5$ \\
\hline
$|\vec{p}_T(\ell_1)-\vec{p}_T(\ell_2)|>150$ GeV \\
\hline
$\cos (\delta\phi_{\ell\ell})<-0.6$ \\
\hline
\end{tabular}
\caption{Additional selection cuts for channel $2j\ell^\pm\ell^\pm\nu\nu$.}
\label{tab:cuts_llss_1}
%\end{table}
}

In the following, as already mentioned, we will consider the full sample as our signal.
It is possible to improve the discriminating power of the analysis increasing the
fraction of $\ordEW$ events in the event sample since only those are sensitive to the
mechanism of EWSB.
Therefore, on the generated samples we have applied the additional selection
cuts shown in \tbn{tab:cuts_llss_1}.
These cuts force the two tag jets to be
well separated and not central.  One of the two leading jets is forbidden from
having a very large transverse momentum. The two charged leptons are required to be rather
central and well separated from the jets. They are required to be
well separated in the transverse plane and to have large transverse momentum.
Finally, the vector difference between the lepton momenta is required to be large. 

\begin{figure}[ht!]
\centering
\subfigure{	 
\hspace*{-2.1cm} 
\includegraphics*[width=8.3cm,height=6.2cm]{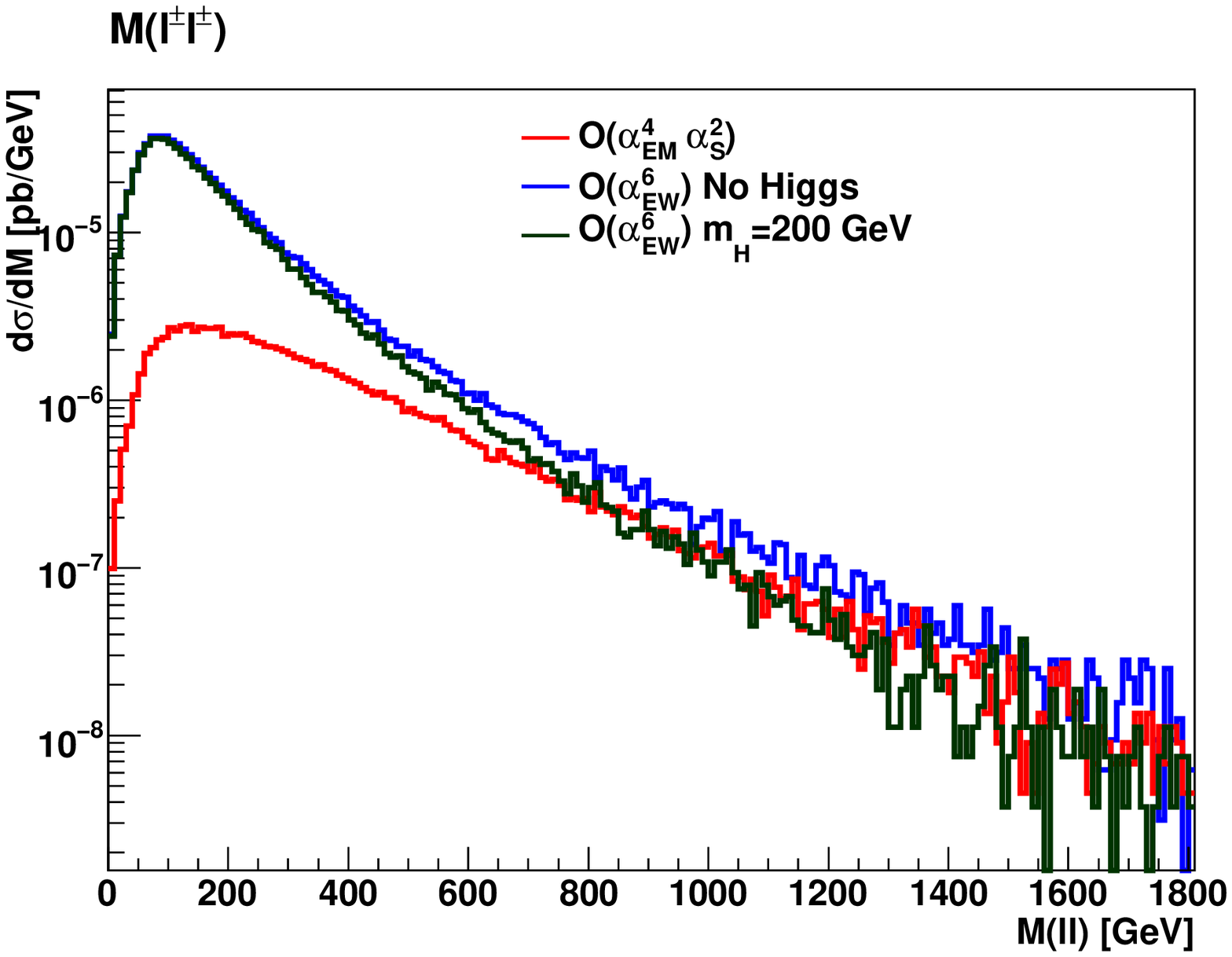}
\hspace*{-0.7cm}
\includegraphics*[width=8.3cm,height=6.2cm]{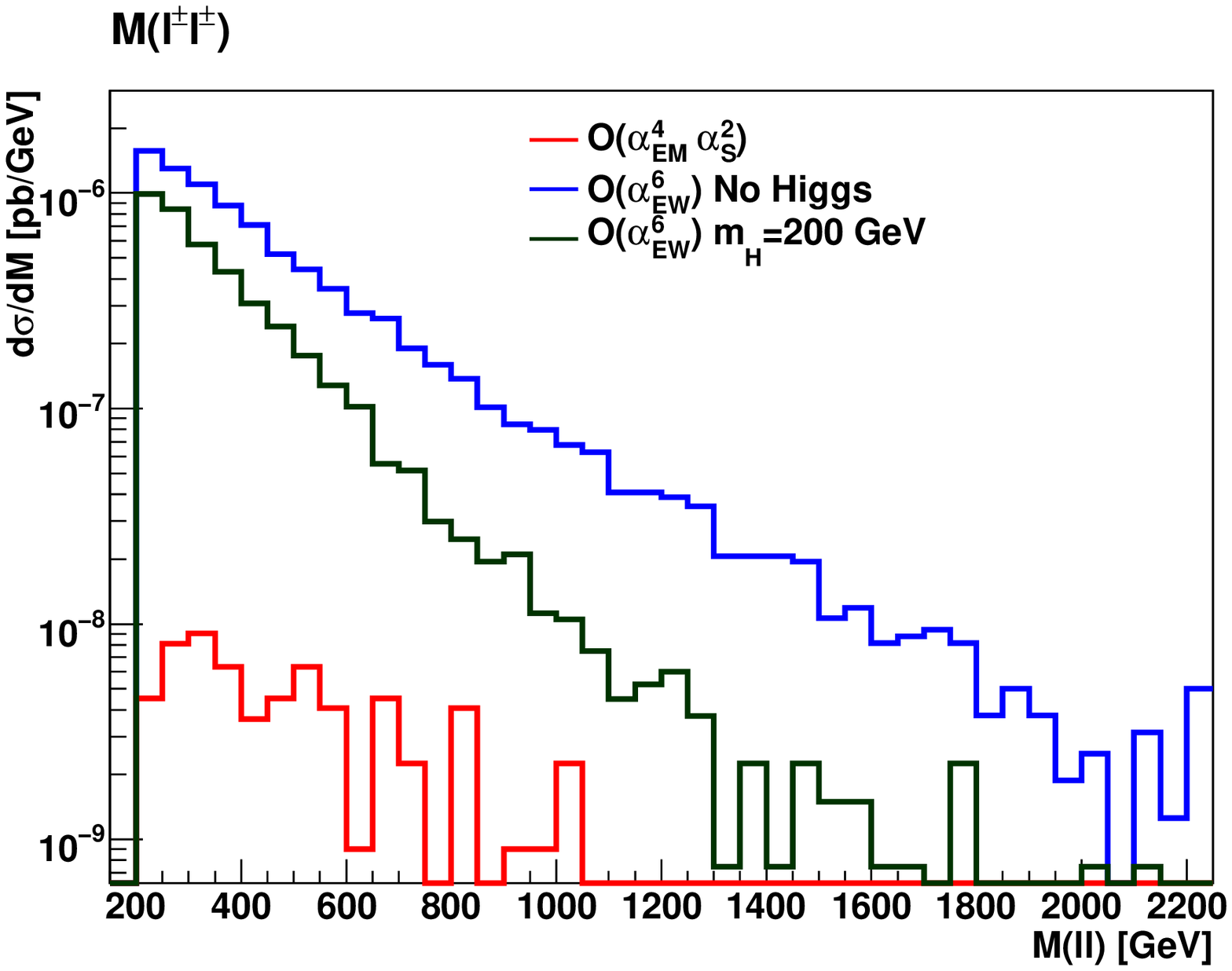}
\hspace*{-3cm}
}
\caption{$M(\ell\ell)$ distribution with acceptance cuts only, \tbn{tab:cuts_0} (left) 
and after all additional selection cuts, \tbn{tab:cuts_llss_1} (right).}
\label{fig:llss_Mlltsum}
\end{figure}

The total cross section in
femtobarns for the $2j\ell^\pm\ell^\pm\nu\nu$ channel, with the full set
of cuts in \tbn{tab:cuts_0}
and \tbn{tab:cuts_llss_1}, as a function of the minimum invariant mass
$M_{cut}$ is shown in
\tbn{tab:res_llss_1}. In parentheses the results for the $\ordEW$ contribution,
which dominate the cross section, are reported.
The distribution of $M(\ell\ell)$ is presented on the right hand side of 
\fig{fig:llss_Mlltsum} which clearly demonstrates the good separation between
the two scenarios obtained through the additional cuts. As expected the separation
increases with increasing di-lepton invariant mass.

%\begin{table}[ht!]
\TABLE{
\begin{tabular}{|p{0.08\textwidth}|p{0.2\textwidth}|p{0.1\textwidth}|p{0.2\textwidth}|p{0.1\textwidth}|p{0.2\textwidth}|}
\hline
$M_{cut}$ 	& \multicolumn{2}{|c|}{no Higgs}& \multicolumn{2}{|c|}{SILH} 	& $M_H=200$ GeV   \\ 
		\cline{2-6}
 (GeV) 		& \centering $\sigma$(fb) & \centering PBSM & \centering $\sigma$(fb) & \centering PBSM	& 
                                                                                
\parbox[t]{0.2\textwidth}{\centering $\sigma$(fb)} \\
\hline
200		&.435(.431)   & 94.9\% & .276 (.273) & 39.1\% & .206(.203)   \\ 
300		&.290(.288)   & 98.2\% & .166 (.164) & 42.3\% & .114(.111)   \\ 
\hline
400		&.191(.189)   & 98.7\% & .0977(.0958)& 41.2\% & .0629(.0609)  \\ 
\hline
500		&.129(.128)   & 98.7\% & .0604(.0588)& 34.4\% & .0351(.0336)  \\ 
600		&.0886(.0876) & 97.5\% & .0385(.0375)& 37.1\% & .0194(.0183)  \\ 
700		&.0614(.0607) & 96.6\% & .0262(.0254)& 42.3\% & .0112(.0105)  \\ 
800		&.0438(.0432) & 91.1\% & .0184(.0178)& 31.2\% & .00701(.00640) \\ 

\hline
\end{tabular}
\caption{
Total cross section for the $\ell^\pm\ell^\pm\nu\nu+2j$ channel in femtobarns,
with the full set of cuts in \tbn{tab:cuts_0} and 
\tbn{tab:cuts_llss_1}, as a function of the minimum
dilepton invariant mass $M_{cut}$ for the 
$\ell^\pm\ell^\pm$ system. In parentheses the results for the 
$\ordEW$ 
sample. We also show the PBSM probabilities. The result for $M_{cut} = 400\mbox{ GeV}$
which provides the best discrimination between the Higgsless and light Higgs
scenario is highlighted.}
\label{tab:res_llss_1}
%\end{table}
}

\begin{figure}[ht!]
\centering
\includegraphics*[width=8.3cm,height=6.2cm]{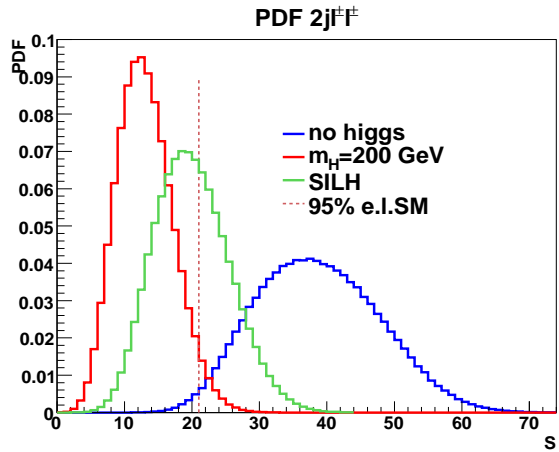}
\caption{Probability distribution for no--Higgs, SILH and SM cases for the
$2j\ell^\pm\ell^\pm\nu\nu$ channel.
The vertical line indicates the 95\%CL for the SM. $M_{cut} = 400 \mbox{ GeV}$.}
\label{fig:prob_llss}
\end{figure}

In \tbn{tab:res_llss_1} we also give the PBSM@95\%CL for the two BSM scenarios.
The corresponding normalized frequency for the three scenarios,
as a function of the number of events, is reported in \fig{fig:prob_llss}
for $M_{cut} = 400 \mbox{ GeV}$. The red curve refers to the probability
distribution for a Higgs of 200 GeV
while the green one refers to the SILH model and the blue one to the no--Higgs
case. The dotted vertical line in the plot marks the 95\% exclusion limit for
the SM predictions.

The probability of an experiment to find a result
incompatible with the SM at 95\%CL, assuming that the Higgsless model is realized
in Nature, is of the order of 99\%  for $M_{cut} = 400\mbox{ GeV}$
and decreases to about 90\% for $M_{cut} = 800
\mbox{ GeV}$. Because of the absence of large backgrounds
this channel has a discriminating power which is in fact quite high.
The corresponding probabilities for
the SILH model vary between 30\% and 40\%.

About 40(20) events are predicted for the Higgsless(SILH) scenario assuming
our standard luminosity of $L=200 \ifb$
and $M_{cut} = 400\mbox{ GeV}$, which provides the best discrimination between the Higgsless
scenario and the SM.

%%%%%%%%%%%%%%%%%%%%%%%%%%%%%%%%%%%%%%%%%%%%%%%%%%%%%%%%%%%%%%%%%%%%%%%%%
\section{The $2jZZ \rightarrow 2j\ell^+\ell^-\nu\bar{\nu}$ channel:
two opposite--sign same--flavour leptons in the final state}
\label{sec:zz}

This channel has been separated from the $2jWW \rightarrow 2j\ell^+\ell^-\nu\bar{\nu}$
case using the di--lepton mass. If $|M(\ell\ell)-M_Z|<15$ GeV, 
the event is considered as produced by a $ZZ$ intermediate state.
Since the mass of the final state $ZZ$ system cannot be fully
reconstructed we estimate the center of mass energy 
of the vector boson scattering from the transverse mass:

\begin{equation}
\label{eqn:mtzz}
M_T^2(ZZ)=[\sqrt{M_Z^2+p_T^2(\ell\ell)}+\sqrt{M_Z^2+p_{Tmiss}^2}]^2-|\vec{p_T}(\ell\ell)+\vec{p}_{Tmiss}|^2
\end{equation}

The total cross section for the $2jZZ \rightarrow 2j\ell^+\ell^-\nu\bar{\nu}$
channel with the acceptance cuts
in \tbn{tab:cuts_0} is presented in \tbn{tab:res_zz_0} as a function of the minimum 
$M_{cut} = M_T(ZZ)$. In parentheses the results for the $\ordEW$
processes. 

The $M_T(ZZ)$ distribution, with acceptance cuts only, is presented on the left hand side
of \fig{fig:zz_MZZtsum}.
The QCD background is much larger than in the channels discussed in \sect{sec:llss}.
The contribution from top pair production is large, particularly at small transverse masses,
even though we are requiring a lepton pair with an invariant mass in the neighborhood of the
$Z$ mass. This contribution rapidly fades at large $M_T(ZZ)$ where the QCD processes without
top are dominating. Moreover, since no requirement of large missing transverse momentum
has been imposed, additional backgrounds at small transverse masses are generated by
$2jZ\rightarrow 2j\ell^+\ell^-$ production. We have not included this
background contribution in our
study since large $p_{Tmiss}$ is demanded in our final analysis and this additional
contribution is completely eliminated.

%\begin{table}[ht!]
\TABLE{
\begin{tabular}{|p{0.08\textwidth}|p{0.27\textwidth}|p{0.27\textwidth}|p{0.26\textwidth}|}
\hline
$M_{cut}$ 	& \centering no Higgs 	& \centering SILH  & 
					\parbox[t]{0.26\textwidth}{\centering $M_H=200$ GeV}   \\ 
		\cline{2-4}
 (GeV) 		& \centering $\sigma$(fb) & \centering $\sigma$(fb) & \parbox[t]{0.26\textwidth}{\centering $\sigma$(fb)} \\
\hline
300             & 1.84 (.607)  & 1.73 (.494)  & 1.70(.464)  \\
400             & .675 (.319)  & .578 (.222)  & .544(.187)  \\
500             & .363 (.197)  & .288 (.122)  & .262(.0962)  \\
600             & .223 (.134)  & .161 (.0727) & .140(.0515) \\
700             & .143 (.0952) & .0947(.0466) &.0781(.0300)  \\
800             & .0926(.0686) & .0553(.0313) &.0426(.0186) \\
900             & .0646(.0515) & .0341(.0210) &.0251(.0120)  \\

\hline
\end{tabular}
\caption{Total cross section for the $(ZZ)\ell^+\ell^-\nu\nu+2j$ channel after generation cuts,
\tbn{tab:cuts_0}.
In parentheses the results for the $\ordEW$ sample. 
}
\label{tab:res_zz_0}
%\end{table}
}

%\begin{table}[ht!]
%\centering
%\vspace{0.5cm}
\TABLE{
\begin{tabular}{|l|}
\hline
$\Delta \eta (jj) > 4.5$ \\
\hline
$M(jj) > 800$ GeV \\
\hline
$\Delta \eta (\ell j)> 1.3$ \\
\hline
$ p_{Tmiss} >  120$ GeV \\
\hline
$|\vec{p}_T(\ell^+\ell^-)-\vec{p}_T^{\,miss}|>290$ GeV \\
\hline
$p_T(\ell^+\ell^-)>120$ GeV \\
\hline
$|\eta(j)|>1.9$ \\
\hline
\end{tabular}
\caption{Selection cuts for channel $(ZZ)\ell^+\ell^-\nu\nu+2j$.}
\label{tab:cuts_zz_1}
%\end{table}
}

\begin{figure}[ht!]
\centering
\subfigure{	 
\hspace*{-2.1cm} 
\includegraphics*[width=8.3cm,height=6.2cm]{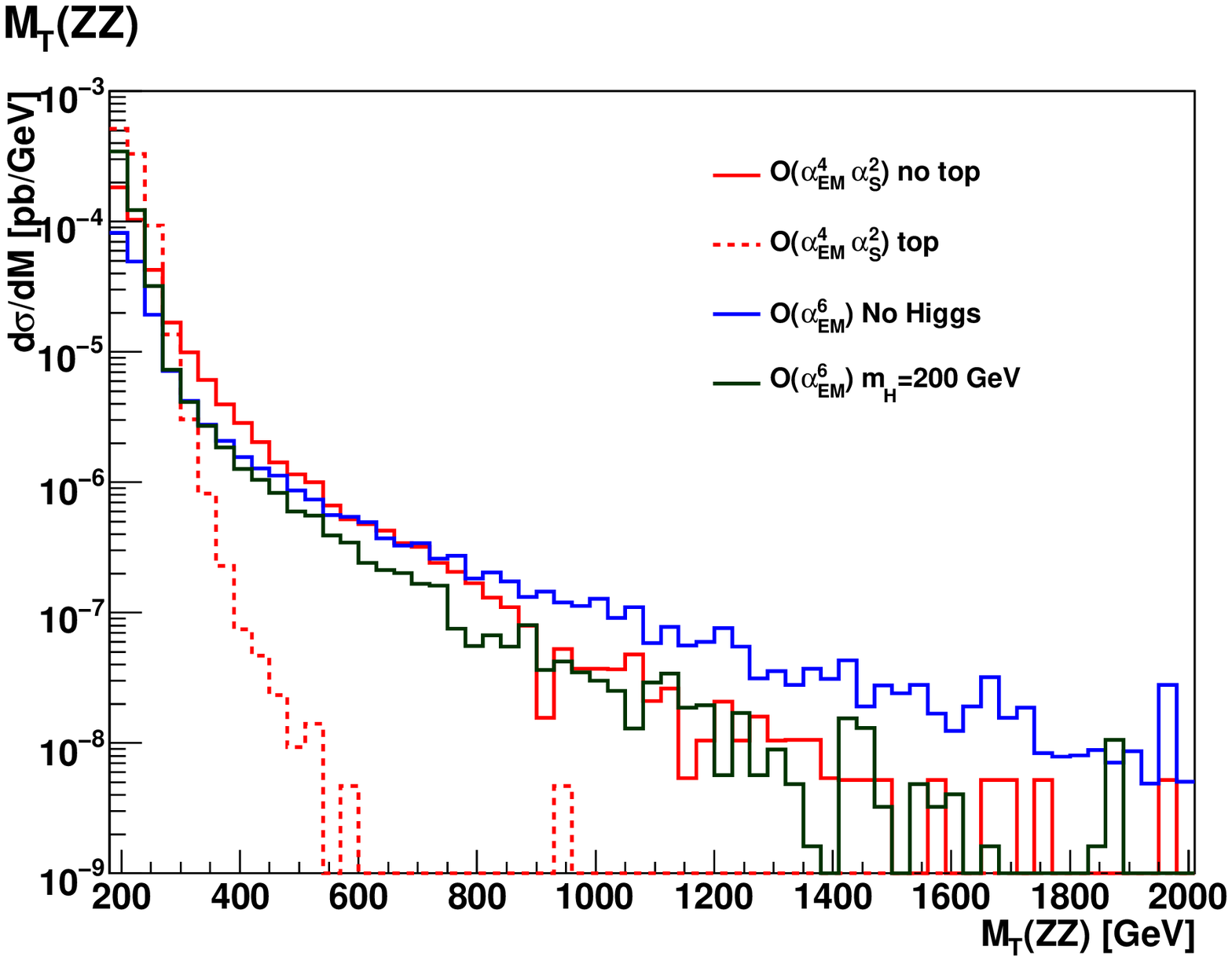}
\hspace*{-0.7cm}
\includegraphics*[width=8.3cm,height=6.2cm]{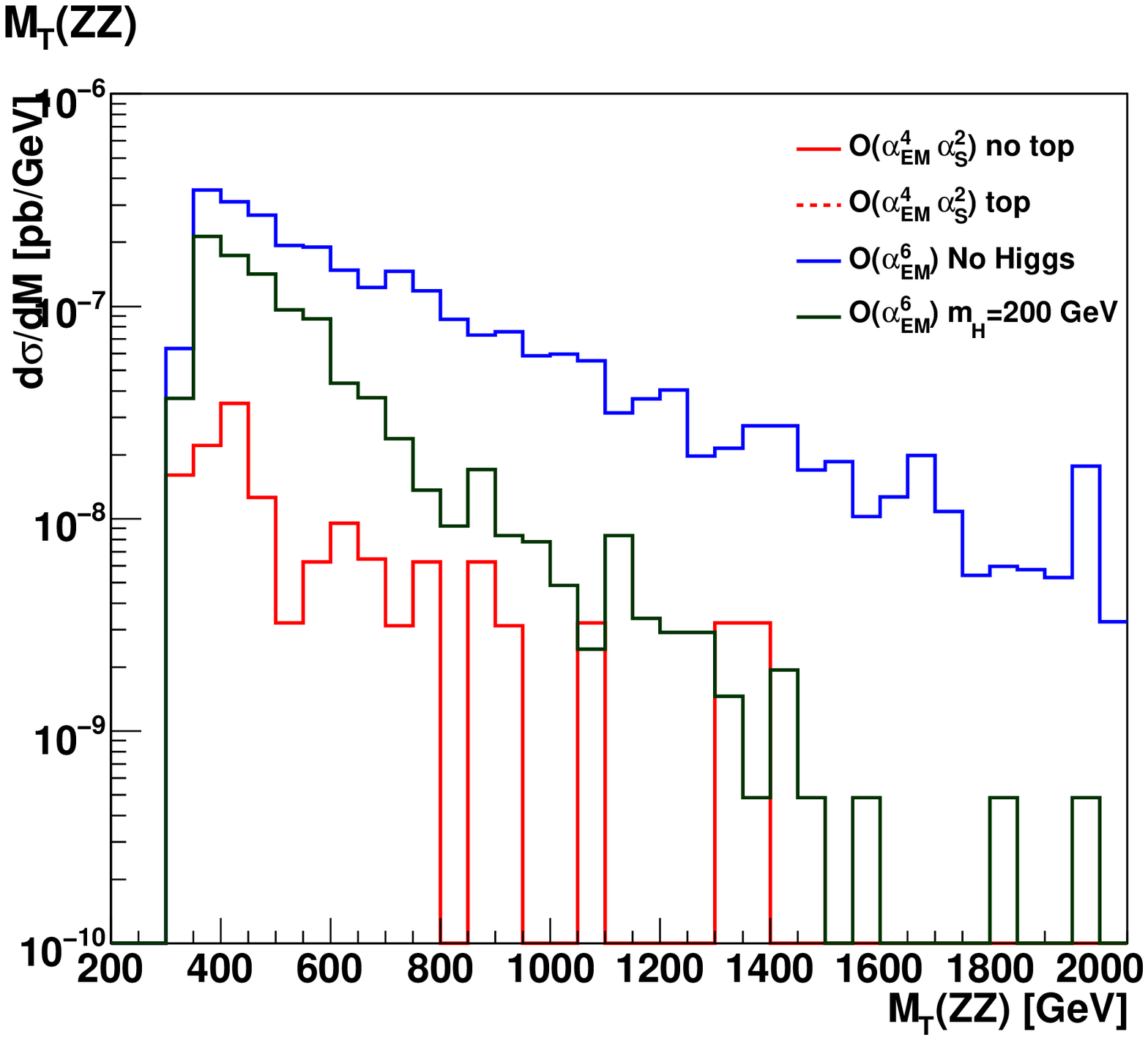}
\hspace*{-3cm}
}
\caption{ZZ transverse mass distribution with initial cuts, \tbn{tab:cuts_0} (left) 
and adding extra cuts, \tbn{tab:cuts_zz_1} (right).}
\label{fig:zz_MZZtsum}
\end{figure}

In order to sharpen the separation between the Standard Model results and those
from alternative symmetry breaking scenarios we impose the additional cuts in 
\tbn{tab:cuts_zz_1}.

The total cross section in
femtobarns for the $2jZZ \rightarrow 2j\ell^+\ell^-\nu\bar{\nu}$ channel,
with the full set of cuts in \tbn{tab:cuts_0}
and \tbn{tab:cuts_zz_1}, as a function of the minimum $ZZ$ transverse mass
$M_{cut}$ is shown in
\tbn{tab:res_zz_1}. In parentheses the results for the $\ordEW$ contribution.
The PBSM probabilities are also given.

%\begin{table}[ht!]
\TABLE{
\begin{tabular}{|p{0.08\textwidth}|p{0.2\textwidth}|p{0.1\textwidth}|p{0.2\textwidth}|p{0.1\textwidth}|p{0.2\textwidth}|}
\hline
$M_{cut}$ 	& \multicolumn{2}{|c|}{no Higgs}& \multicolumn{2}{|c|}{SILH} 	& $M_H=200$ GeV   \\ 
		\cline{2-6}
 (GeV) 		& \centering $\sigma$(fb) & \centering PBSM & \centering $\sigma$(fb) & \centering PBSM	& 
                                                                                
\parbox[t]{0.2\textwidth}{\centering $\sigma$(fb)} \\
\hline
300		&.143(.136)   & 94.6\% & .0770(.0700)  & 31.5\% &.0540(.0470)   \\ 
400		&.120(.115)   & 96.1\% & .0614(.0564)  & 36.2\% &.0396(.0345)   \\ 
500		&.0887(.0860) & 97.5\% & .0396(.0369)  & 39.8\% &.0214(.0187)   \\ 
\hline
600		&.0691(.0668) & 98.4\% & .0268(.0246)  & 44.3\% &.0118(.00957) \\ 
\hline
700		&.0547(.0533) & 97.0\% & .0186(.0171)  & 32.0\% &.00697(.00555) \\ 
800		&.0410(.0401) & 94.6\% & .0145(.0136)  & 33.2\% &.00463(.00368) \\ 
900		&.0327(.0321) & 94.3\% & .00991(.00927)& 31.9\% &.00300(.00236) \\ 
\hline
\end{tabular}
\caption{
Total cross section for the $(ZZ)\ell^+\ell^-\nu\nu+2j$ channel in femtobarns,
with the full set of cuts in \tbn{tab:cuts_0} and 
\tbn{tab:cuts_zz_1}, as a function of the minimum transverse mass $M_T(ZZ)_{cut}$. 
In parentheses the results for the 
$\ordEW$ 
sample. The PBSM@95\%CL are also shown.
}
\label{tab:res_zz_1}
%\end{table}
}

\begin{figure}[ht!]
\centering
\includegraphics*[width=8.3cm,height=6.2cm]{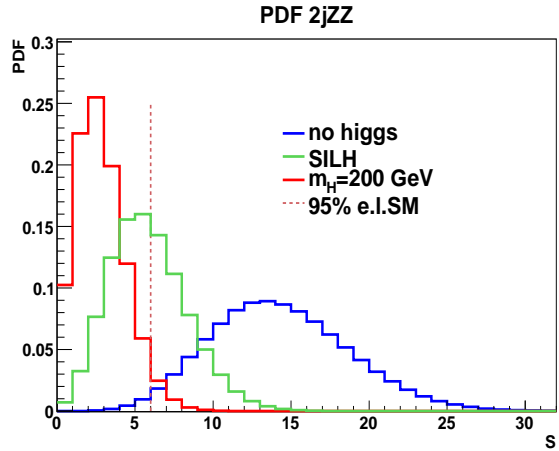}
\caption{Probability distribution for the no--Higgs, SILH and SM cases for the
$2jZZ \rightarrow 2j\ell^+\ell^-\nu\bar{\nu}$ channel.
The vertical line indicates the 95\%CL for the SM. $M_{cut} = 600 \mbox{ GeV}$.}
\label{fig:prob_zz}
\end{figure}

The $M_T(ZZ)$ distribution, with the full set of cuts, is presented on the
right hand side of \fig{fig:zz_MZZtsum}. The $\ordQCD$ background has been sharply
reduced while increasing the separation between the Higgsless and
light Higgs scenarios. 
The top related background has been totally suppressed.

The probability distribution for the three scenarios is reported in
\fig{fig:prob_zz}
for $M_T(ZZ)_{cut} = 600 \mbox{ GeV}$. The red curve refers to a Higgs of 200 GeV
while the green one refers to the SILH model and the blue one to the no--Higgs
case. The dotted vertical line in the plot marks the 95\% exclusion limit for
the SM predictions.
The probability of an experiment to find a result
incompatible with the SM at 95\%CL, assuming that the Higgsless model is realized
in Nature, is of the order of 98\%  for $M_{cut} = 600\mbox{ GeV}$ and does not
vary significantly over the range considered in \tbn{tab:res_zz_1}.
The corresponding probabilities for
the SILH model vary between 30\% and 45\%.

%\eject

%%%%%%%%%%%%%%%%%%%%%%%%%%%%%%%%%%%%%%%%%%%%%%%%%%%%%%%%%%%%%%%%%%%%%%%%%
\section{The $2j W^+W^- \rightarrow 2j\ell^+\ell^{\prime-}\nu\bar{\nu}$ channel:
two opposite--sign leptons in the final state}
\label{sec:ww}

The total cross section for the $2j W^+W^- \rightarrow 2j\ell^+\ell^-\nu\bar{\nu}$
channel, with the acceptance
cuts in \tbn{tab:cuts_0}, is shown in \tbn{tab:res_ww_0} as a function of the
minimum $\ell\ell$ invariant mass. 
As usual, in parentheses we show the results for the $\ordEW$ processes.
The cross sections for $t\bar{t}j$ and $t\bar{t}jj$ production are presented separately.
We have required exactly two jets in the acceptance region. This however is not sufficient
to guarantee a well defined cross section because two $b$ quarks are produced in the
leptonic decay of the tops. Therefore we have further required that one $b$ for
$t\bar{t}j$ events and both $b$'s for $t\bar{t}jj$ ones remain undetected.
We consider a $b$--quark detected if $\vert \eta_b \vert < 5$   and $p_{Tb} > 30$ GeV. 
As a consequence the partons produced in association with the $t\bar{t}$ pair
are forced to be
visible and the corresponding cross section is finite.
The phase space regions which are excluded by these constraints are eliminated by the
cut on the mass of all lepton-jet pairs discussed below.

This process has the largest production rate
among all channels considered in this paper,
however the QCD background is much larger than the electroweak part.

%\begin{table}[ht!]
%\centering
\TABLE{
\begin{tabular}{|c|c|c|c|c|c|}
\hline
$M_{cut}$ 	& no Higgs & SILH  & $M_H=200$ GeV & $t\bar{t}j$ & $t\bar{t}jj$  \\ 
		\cline{2-6}
 (GeV) 		& $\sigma$(fb) & $\sigma$(fb) & $\sigma$(fb) & $\sigma$(fb) & $\sigma$(fb)\\
\hline
300             & 70.0(4.65) & 69.7(4.42)  & 69.7(4.35) & 39.7   & 2.59 \\
400             & 29.7(2.32) & 29.5(2.11)  & 29.5(2.12) & 16.4   & 1.22\\
500             & 13.5(1.24) & 13.4(1.14)  & 13.4(1.13) & 7.21   & .516\\
600             & 6.69(.713) & 6.62(.643)  & 6.60(.627) & 3.13   & .237\\
700             & 3.55(.440) & 3.49(.376)  & 3.47(.362) & 1.45   & .139\\
800             & 1.92(.274) & 1.88(.236)  & 1.87(.225) & .849   & .0737\\
\hline
\end{tabular}
\caption{Total cross section for the $(W^+W^-)\ell^+\ell^-\nu\nu+2j$ channel after initial cuts,
\tbn{tab:cuts_0} in function of the minimum $\ell\ell$ invariant mass, $M(\ell\ell)$.
In parentheses the results for the $\ordEW$ sample. 
}
\label{tab:res_ww_0}
%\end{table}
}

The $M(\ell\ell)$ distribution, with acceptance cuts only, is presented on the left hand side
of \fig{fig:ww_Mllsum}.
$t\bar{t}$ production is very important at this level, 
and the usual way to suppress it, by requiring $M(Wj)$ out of the top nominal
mass window, is not applicable because of the impossibility to reconstruct
the $W$ mass. 
Instead we require the mass of all lepton-jet pairs to be larger than the top mass.

%Another important source of background is represented by 
%EW top-production which contributes equally to the Higgsless and to the light
%Higgs scenarios. Indeed the two results are hardly distinguishable at this stage.

The relatively high signal rate and the large background allow and require
harder cuts than in all previous cases. 
The additional selection requirements are shown in \tbn{tab:cuts_ww_1}.
The constraint on the lepton-jet mass
is quite effective in reducing the background due to top pair production.
The $t\bar{t}$ and $t\bar{t}j$ contributions are essentially eliminated and the two light partons in 
$t\bar{t}jj$ production are forced to be tagged.
However this cut reduces significantly the boson scattering signal and furthermore
it increases the relative contribution of $t\bar{t}jj$.

%\begin{table}[tbh!]
%\centering
\TABLE{
\begin{tabular}{|l|}
\hline
$M(jj) > 1000$ GeV \\
\hline
$\Delta \eta (jj) > 4.8$ \\
\hline
$|\eta(\ell)|< 2.00$ \\
\hline
$p_T(\ell) >  40$ GeV \\
\hline
$max|\eta(j)|> 2.5$ \\
\hline
$|\eta(j)|> 1.3$ \\
\hline
$E(j)> 180$ GeV \\
\hline
$\Delta \eta (\ell j)> 0.8$ \\
(and $\Delta R (\ell j)> 1$) \\
\hline
$M(\ell j)>180$ GeV \\
\hline
$|\vec{p}_T(\ell^+)-\vec{p}_T(\ell^-)|>220$ GeV \\
\hline
$\cos (\delta\phi_{\ell\ell})<-0.6$ \\
\hline
\end{tabular}
\caption{Additional selection cuts for the $2jW^+W^- \rightarrow 2j\ell^+\ell^-\nu\nu$ channel.}
\label{tab:cuts_ww_1}
%\end{table}
}

The total cross section in
femtobarns for the $(W^+W^-)\ell^+\ell^-\nu\nu+2j$ channel,
with the full set of cuts in \tbn{tab:cuts_0}
and \tbn{tab:cuts_ww_1}, as a function of the minimum di--lepton invariant mass
$M_{cut}$ is shown in
\tbn{tab:res_ww_1}. In parentheses the results for the $\ordEW$ contribution.
The PBSM probabilities are also presented. As mentioned in \sect{sec:calculation}
only the statistical uncertainty has been taken into account for the $t\bar{t}jj$
background. 
The requirement that both $b$--quarks have a transverse momentum smaller
than 30 GeV or a rapidity in modulus larger than 5 units is quite stringent, in fact the cross section
for $t\bar{t}jj$ requiring both $b$'s to produce visible jets, not necessarily identified as $b$--jets,
is at least an order of magnitude larger than the results presented in \tbn{tab:res_ww_1}.
Therefore, we believe the $t\bar{t}jj$ will be measured in the complementary region and extrapolated
to the signal domain with small uncertainty.
In any case, since the hadronic activity is expected to be much higher in  $t\bar{t}$+jets events
than in boson boson scattering ones, a more accurate assessment of this background would require
complete showering and hadronization.

We will come back to the impact on our results in case the $t\bar{t}jj$ could not be measured in
\sect{sec:combining}. For the time being we only report that the PBSM@95\%CL reported in \tbn{tab:res_ww_1}
for $M_{cut}$ = 600 GeV would change from 85\% to 78\% for the no Higgs case and from 27\% to 22\%
for the SILH model.

\begin{figure}[t!]
\centering
\subfigure{	 
\hspace*{-2.1cm} 
\includegraphics*[width=8.3cm,height=6.2cm]{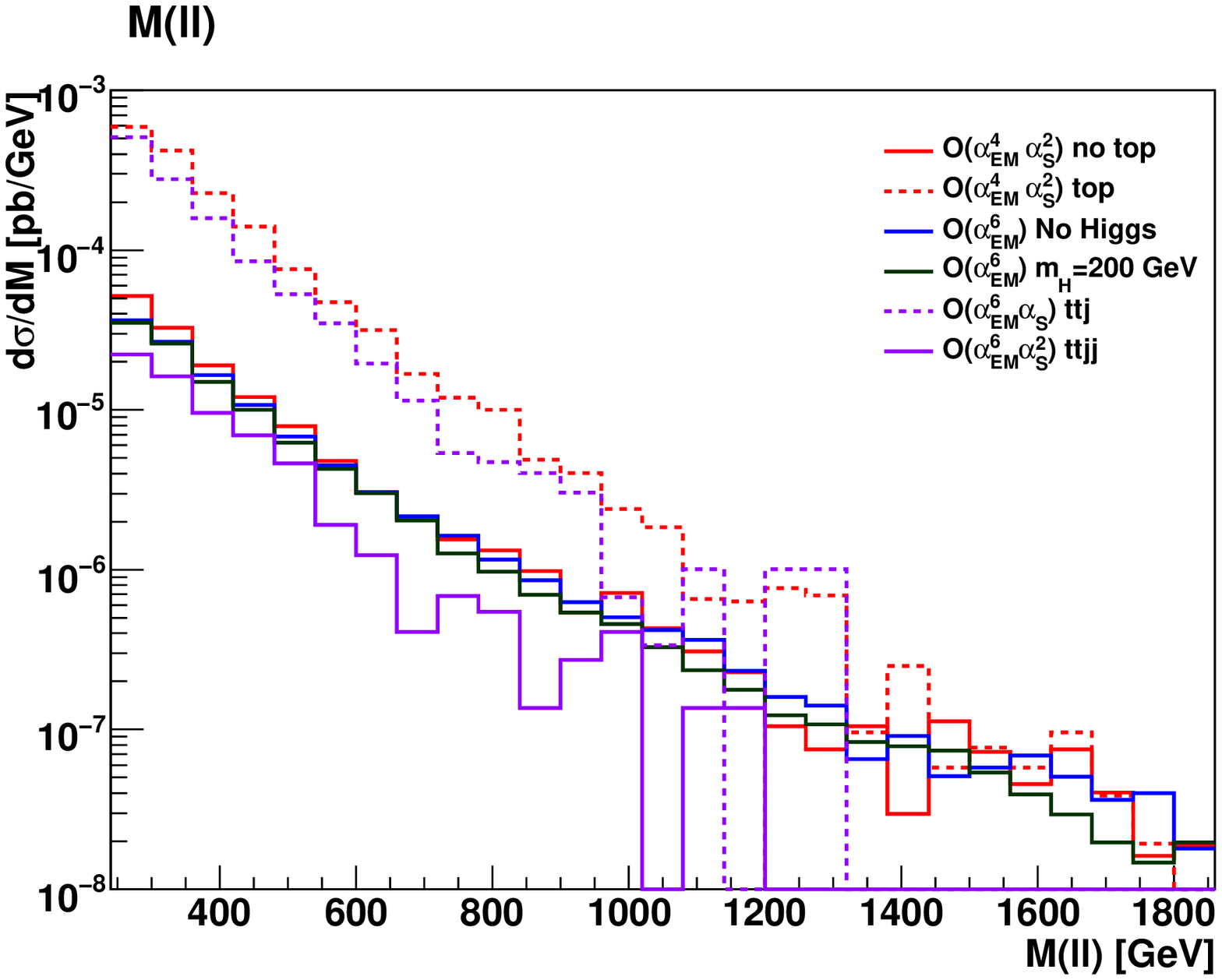}
\hspace*{-0.7cm}
\includegraphics*[width=8.3cm,height=6.2cm]{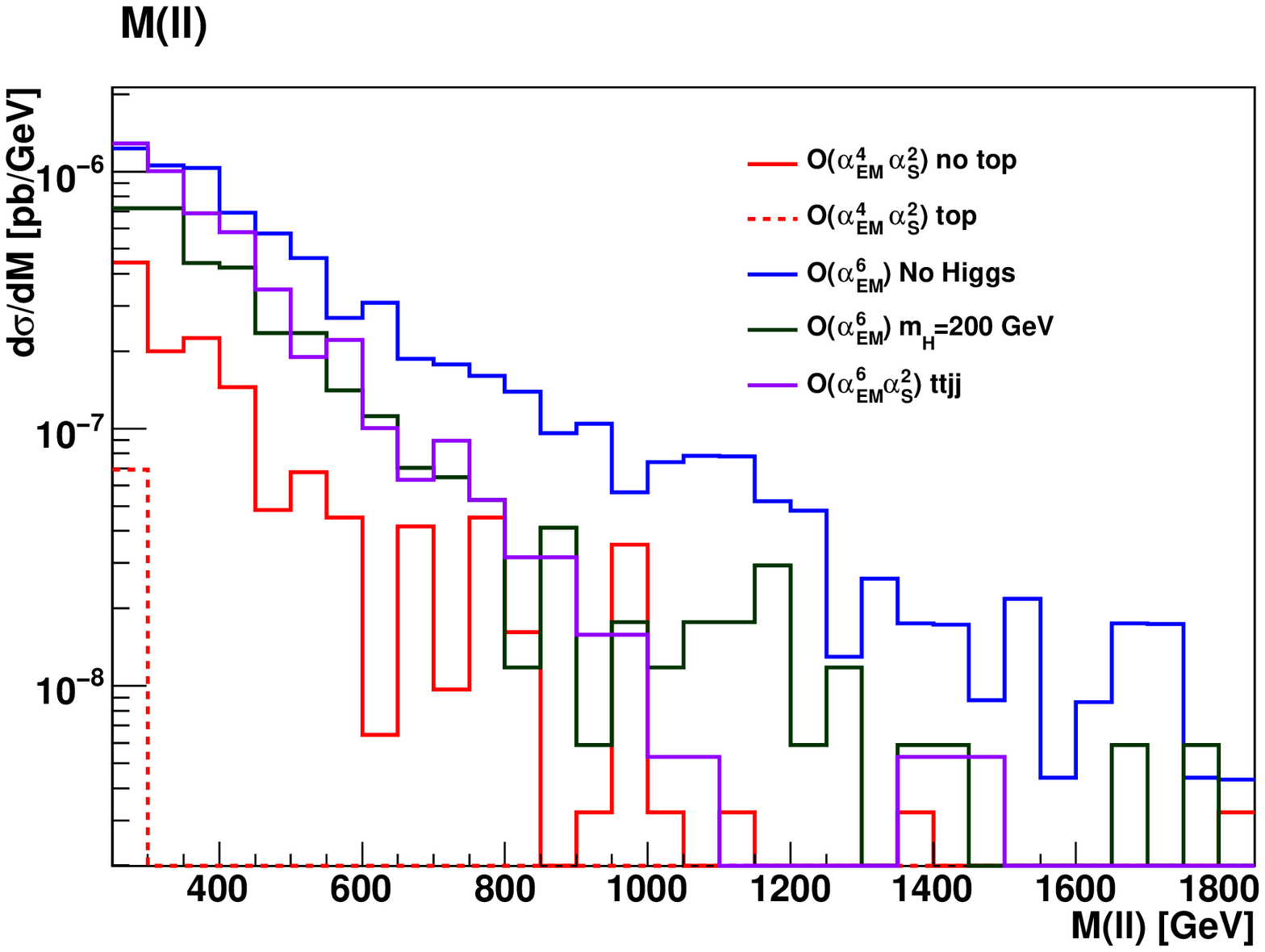}
\hspace*{-3cm}
}
\caption{Di-lepton mass distribution with initial cuts, \tbn{tab:cuts_0} (left) 
and adding extra cuts, \tbn{tab:cuts_ww_1} (right).}
\label{fig:ww_Mllsum}
\end{figure}

%\begin{table}[ht!]
%\centering
\TABLE{
\begin{tabular}{|c|c|c|c|c|c|c|}
\hline
$M_{cut}$ 	& \multicolumn{2}{|c|}{no Higgs}& \multicolumn{2}{|c|}{SILH} 	&  $M_H=200$ & $t\bar{t}jj$\\ 
		\cline{2-7}
 (GeV) 		& \centering $\sigma$(fb) & \centering PBSM & \centering $\sigma$(fb) & \centering PBSM	& 
\centering $\sigma$(fb) & $\sigma$(fb)\\
\hline
300		& .337(.292)   & 79.58\% & .224(.179)   & 22.69\% & .179(.134)   & .173\\
400		& .212(.188)   & 80.74\% & .131(.107)   & 20.89\% & .100 (.0765) & .0890\\
500		& .139(.125)   & 82.83\% & .0841(.0700) & 26.35\% & .0577(.0435) & .0407\\
\hline
600		& .0968(.0883) & 85.03\% & .0533(.0448) & 26.56\% & .0332(.0247) & .0215\\
\hline
700		& .0696(.0635) & 80.55\% & .0353(.0292) & 20.63\% & .0217(.0156) & .0138\\
\hline
\end{tabular}
\caption{
Total cross section for the $2jW^+W^- \rightarrow 2j\ell^+\ell^-\nu\nu$ channel in femtobarns,
with the full set of cuts in \tbn{tab:cuts_0} and \tbn{tab:cuts_ww_1},
as a function of the minimum di-lepton invariant mass $M(\ell\ell)_{cut}$. 
In parentheses the results for the 
$\ordEW$ 
sample. The PBSM@95\%CL are given in the third and fifth column.
}
\label{tab:res_ww_1}
%\end{table}
}

\begin{figure}[ht!]
\centering
\includegraphics*[width=8.3cm,height=6.2cm]{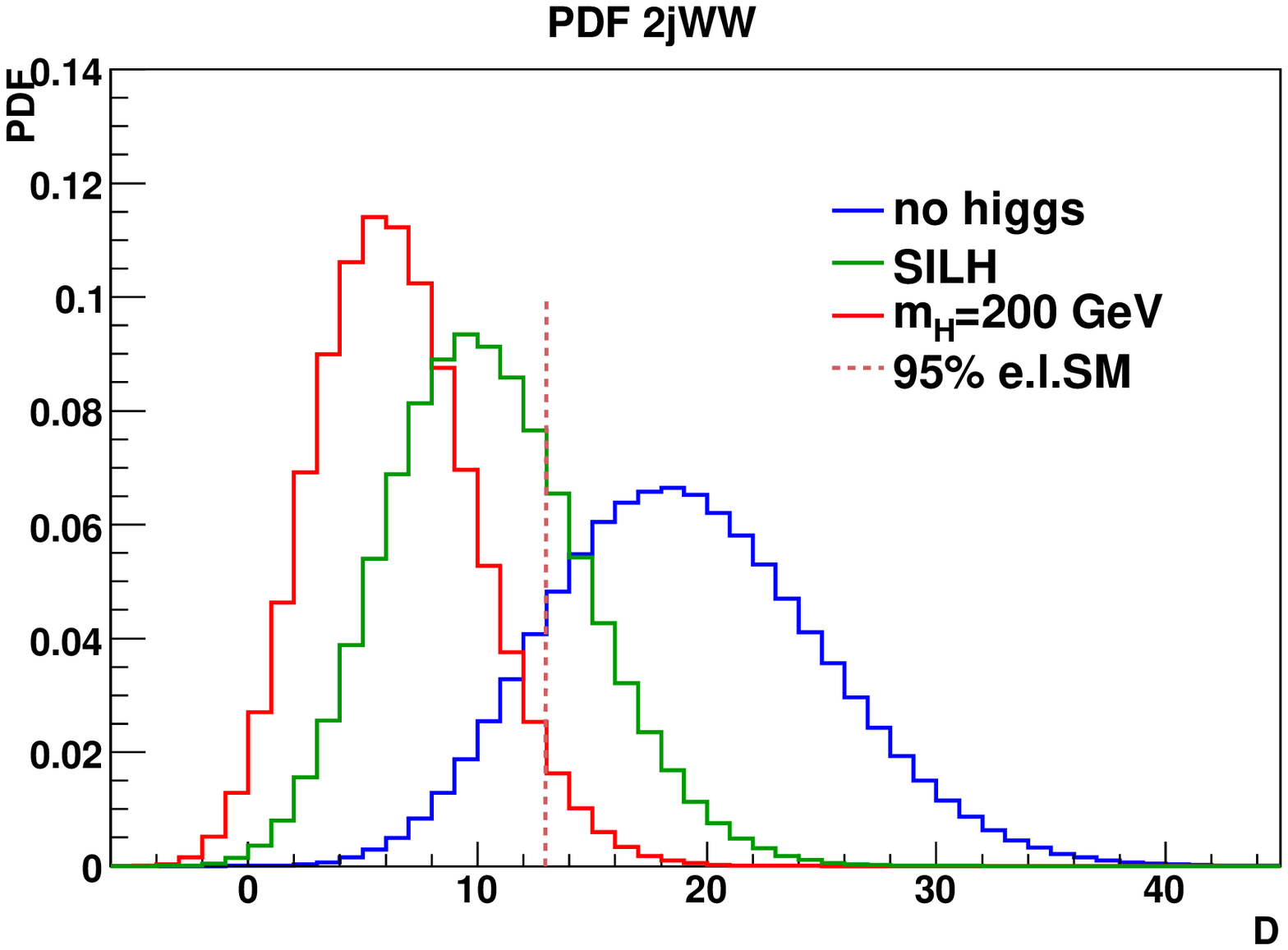}
\caption{Probability distribution for the no--Higgs, SILH and SM cases for the 
$2jW^+W^- \rightarrow 2j\ell^+\ell^-\nu\bar{\nu}$ channel.
The vertical line indicates the 95\%CL for the SM. $M_{cut} = 600 \mbox{ GeV}$.}
\label{fig:prob_ww}
\end{figure}

The di-lepton mass distribution, with the full set of cuts, is presented on the
right hand side of \fig{fig:ww_Mllsum}. The $\ordQCD$ background is now
very small while the separation between the Higgsless and light Higgs scenarios is
clearly visible.

The probability distribution for the three scenarios is reported in
\fig{fig:prob_ww} for $M(ll)_{cut} = 600 \mbox{ GeV}$.
The probability of an experiment to find a result
incompatible with the SM at 95\%CL, assuming that the Higgsless model is realized
in Nature, is between 80 and 85\%  for
$300 \mbox{ GeV} < M_{cut} < 700\mbox{ GeV}$.
For the SILH model the corresponding probabilities lie between 20 and 26\%.

%\eject

%%%%%%%%%%%%%%%%%%%%%%%%%%%%%%%%%%%%%%%%%%%%%%%%%%%%%%%%%%%%%%%%%%%%%%%%%
\section{The $2j4\ell$ channel}
\label{sec:4l}

Contrary to all reaction discussed above, in the $2j4\ell$ channel the mass of the
final state vector boson pair can be directly measured to a high precision.
It is presented here for completeness despite its small rate and statistical
discriminating power.

%\begin{table}[ht!]
%\centering
%\vspace{0.5cm}
\TABLE{
\begin{tabular}{|l|}
\hline
$M(jj) > 800$ GeV \\
\hline
$p_T(Z) >  100$ GeV \\
\hline  
$\Delta R (Z j)> 1$ \\
\hline
$\cos (\delta\phi_{ZZ})<-0.4$ \\
\hline
\end{tabular}
\caption{Additional selection cuts for the $2j4\ell$ channel.}
\label{tab:cuts_4l_1}
%\end{table}
}

The QCD contribution is small already at generation level. 
However, for a luminosity of $L = 200 \mbox{ fb}^{-1}$
the difference between the number of events expected for an
infinite mass Higgs and a light one is of the order the statistical
uncertainty for the $\ordQCD$ contribution and no meaningful separation between
the two cases can be obtained.
Only a minimum set of additional cuts
can be applied in order to have  at least a handful of events for $L=200 \ifb$.
This channel could clearly profit from higher luminosities.
The additional selection
cuts are shown in \tbn{tab:cuts_4l_1}. 

%\begin{table}[ht!]
%\centering
\TABLE{
\begin{tabular}{|p{0.08\textwidth}|p{0.2\textwidth}|p{0.1\textwidth}|p{0.2\textwidth}|p{0.1\textwidth}|p{0.2\textwidth}|}
\hline
$M_{cut}$ 	& \multicolumn{2}{|c|}{no Higgs}& \multicolumn{2}{|c|}{SILH} 	& $M_H=200$ GeV   \\ 
		\cline{2-6}
 (GeV) 		& \centering $\sigma$(ab) & \centering PBSM & \centering $\sigma$(ab) & \centering PBSM	& 
                                                                                 \parbox[t]{0.24\textwidth}{\centering $\sigma$(ab)} \\
\hline
300		& 51.8(41.6) & 35.6\%  	& 36.1(26.0)  &	8.4\%	& 31.6(21.5) \\
400		& 44.7(36.7) & 40.7\%	& 30.1(22.1)  &	10.3\%	& 25.5(17.5) \\
\hline
500             & 35.6(30.1) & 41.8\% 	& 22.8(17.3)  & 10.5\% 	& 18.4(12.9)	\\
\hline
600             & 28.2(24.2) & 34.1\%	& 17.2(13.2)  & 7.0\%	& 13.5(9.45)	\\
700             & 22.2(19.5) & 29.3\%	& 12.8(10.0)  & 5.3\%	& 9.64(6.93)	\\
800             & 17.8(15.8) & 29.1\%	& 9.79(7.82)  & 5.4\%   & 7.09(5.12)	\\
900             & 14.0(12.6) & 31.0\%	& 7.38(6.05)  & 6.7\%   & 5.19(3.87)	\\
\hline
\end{tabular}
\caption{
Total cross section for the $4\ell + 2j$ channel in attobarns,
with the full set of cuts in \tbn{tab:cuts_0} and 
\tbn{tab:cuts_4l_1}, as a function of the minimum invariant mass $M_{cut}$ for the 
$4\ell$ system. In parentheses the results for the 
$\ordEW$ 
sample. Also shown are the PBSM probabilities.
}
\label{tab:res_4l_1}
%\end{table}
}

\begin{figure}[ht!]
\centering
\includegraphics*[width=8.3cm,height=6.2cm]{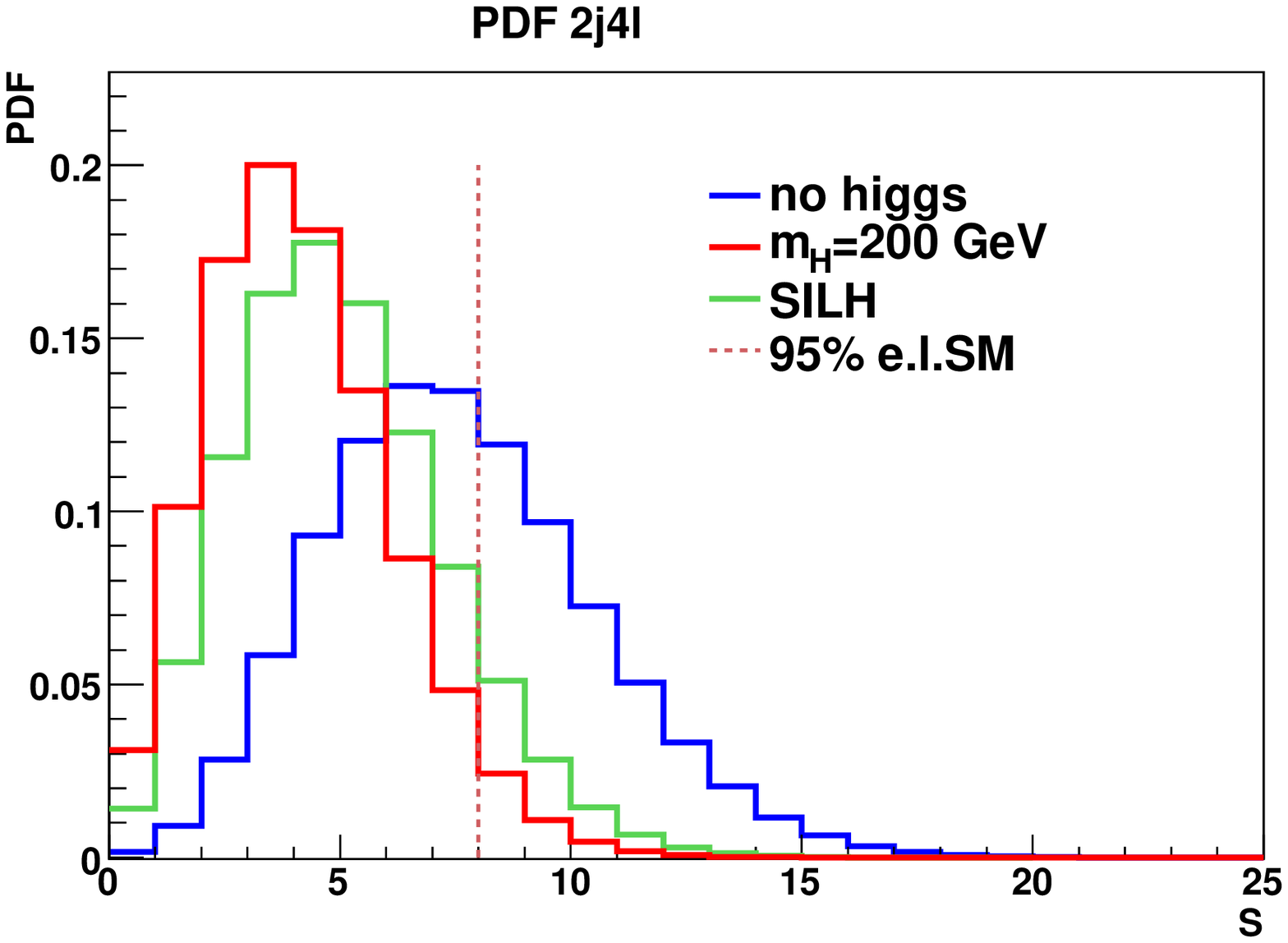}
\caption{Probability distribution for the no--Higgs, SILH and SM cases for the $2j4\ell$ channel.
The vertical line
indicates the  95\%CL for the SM. $M_{cut} = 500 \mbox{ GeV}$.}
\label{fig:prob_4l}
\end{figure}

The cross section after these extra cuts is presented in \tbn{tab:res_4l_1}
as a function of the minimum 
$ZZ$ mass. In parentheses the results for the $\ordEW$ sample.
We also show the PBSM@95\%CL for the standard $L=200 \ifb$ luminosity.

The probability distribution of the discriminant $S$
for the three scenarios is reported in
\fig{fig:prob_4l} for $M_{cut} = 500 \mbox{ GeV}$
with the full set of cuts in \tbn{tab:cuts_0} and \tbn{tab:cuts_4l_1}.

As reported in \tbn{tab:res_4l_1},
the probability of an experiment to find a result incompatible with the SM at
95\%CL, assuming that the Higgsless model is realized in Nature, is of the order
of 42\% for $M_{cut} = 500 \mbox{ GeV}$ and decreases to about 30\% for $M_{cut}
= 700 \mbox{ GeV}$.
For the SILH model the
PBSM@95\%CL is only about 10\% at most, for $M_{cut} = 500 \mbox{ GeV}$.

For $L = 200 \mbox{ fb}^{-1}$ and summing over all final states, the
expected total rates are of the order of 4$\div$8 events for the Higgsless case.
%\eject

%%%%%%%%%%%%%%%%%%%%%%%%%%%%%%%%%%%%%%%%%%%%%%%%%%%%%%%%%%%%%%%%%%%%%%%%%
\section{Combining all channels}
\label{sec:combining}

In this section we derive the probability that, assuming that either the
Higgsless scenario or the instance of SILH model we have considered is realized
in Nature, the results of the measurements of the seven channels 
$2j\ell^\pm\ell^\pm\nu\nu$, $2jZZ\rightarrow 2j\ell\nu\ell\nu$,
$2jWW\rightarrow 2j\ell\nu\ell\nu$,$2j4\ell$,
$4j\ell\nu$,$4j\ell\ell$ and $2j3\ell\nu$
at the LHC yield results which are outside the 95\% probability region for the SM.

\tbn{tab:evrec} shows the number of expected events
for the reactions in which the $VV$ mass can be reconstructed.
The data for $4j\ell\nu$ production
are taken from \rf{Ballestrero:2008gf} while
those for $4j\ell^+\ell^-$ and $2j\ell^+\ell^-\ell'\nu$ are from \rf{Ballestrero:2009vw}
to which we refer for more details.
\tbn{tab:evnonrec} instead shows the number of expected events for the
channels in which direct reconstruction of the vector boson pair mass is impossible.
In both instances the assumed integrated luminosity
is $200\ifb$. In each case
the prediction corresponds to the value of $M_{cut}$  
which gives the best PBSM@95\%CL.  These values are highlighted in Tables
\ref{tab:res_llss_1}, \ref{tab:res_zz_1}, \ref{tab:res_ww_1} and \ref{tab:res_4l_1}
for the reactions described in detail in this paper. For the remaining channels
$M_{cut}$ has been set to $600 \mbox{ GeV}$.

For a given number of events for each channel
$k_1$, $k_2$, ..., $k_n$, with corresponding
mean values $\lambda_1$, $\lambda_2$, ..., $\lambda_n$,
which we will refer to collectively as $\vec{k}$ and $\vec{\lambda}$,
the standard likelihood ratio in \eqn{eq:ratio} can be expressed as
$Q(\vec{k};\vec{\lambda}_{BSM},\vec{\lambda}_{SM})=P(\vec{k},\vec{\lambda}_{BSM})/P(\vec{k},\vec{\lambda}_{SM})$ 

The procedure we have employed so far and in \rf{Ballestrero:2009vw}
to evaluate the PBSM
becomes cumbersome when too many channels have to be considered and the dimensionality
of the integrals in  \eqnsc{eq:alphaFix}{eq:PBSM} becomes large.
Therefore for the combination of all results we have resorted to the variable
$-2 \ln Q$. 
From the one dimensional probability distribution of $-2 \ln Q$ the 95\%CL and 99.7\%CL
limits for the SM can be easily determined.

In the following we will first combine separately
the first three channels, in which the
invariant mass of the $VV$ pair cannot be reconstructed, and the last four,
in which the $VV$ mass can be directly measured.
Later we will proceed to a full combination.

%\begin{table}[ht!]
%\centering
\TABLE{
\begin{tabular}{|l|c|c|c|c|c|}
\hline  		       
			       		  & S(noHiggs)  & S(SILH)   & S($m_H=200$ GeV)  & B $\ordQCDtwo$ & $t\bar{t}jj$ \\
\hline
$4j\ell\nu$		   		  & 473.6	& 281.6		& 210.4		& 1956. & 92.6\\
\hline                                                    
$4j\ell^+\ell^-$	     		  & 61.6 	& 30.4		& 19.38		& 220.	& -- \\
\hline                                                    
$2j\ell^+\ell^-\ell'\nu$		  & 10.8	& 5.4		& 3.4		& --    & -- \\
\hline
$2j\ell^+\ell^-\ell^+\ell^-$   		  & 7.12  	& 4.56	       	& 3.68 		& --    & --\\
\hline
\end{tabular}
\caption{Number of events expected for \textbf{$L=200fb^{-1}$} for the channels
in which the $VV$ mass can be reconstructed.
The cuts for the first three reactions are described in \cite{Ballestrero:2008gf}
and \cite{Ballestrero:2009vw}.
The cuts for the $2j4\ell$ channel are discussed in \sect{sec:4l}.
$M_{cut}$ is chosen in such a way that the best PBSM@95\%CL for each channel is obtained.}
\label{tab:evrec}
%\end{table}
}

%\begin{table}[ht!]
%\centering
\TABLE{
\begin{tabular}{|l|c|c|c|c|}
\hline  		       
			       		  & S(noHiggs)  & S(SILH)       & S($m_H=200$ GeV) & $t\bar{t}jj$ \\
\hline
$2j\ell^\pm\ell^\pm\nu\nu$     		  & 38.2       	& 19.54	       	& 12.58  	   & -- \\
\hline
$2jZZ \rightarrow \ell^+\ell^-\nu\nu$     & 13.82 	& 5.36	       	& 2.36 		   & --\\
\hline
$2jW^+W^- \rightarrow \ell^+\ell^-\nu\nu$ & 19.36        & 10.66	& 6.64 	   	   & 4.3 \\
\hline
\end{tabular}
\caption{Number of events expected for \textbf{$L=200fb^{-1}$} for the channels
in which the $VV$ mass cannot be reconstructed.
$M_{cut}$ is chosen in such a way that the best PBSM@95\%CL for each channel is obtained.}
\label{tab:evnonrec}
%\end{table}
}

The probability $P(\vec{k},\vec{\lambda})$
depends on the correlations between channels. In our simplified approach
in which only statistical and theoretical errors are accounted for, only the 
uncertainties which are related to theory are correlated. Statistical errors
in each channel are independent.

As a first step, we assume each channel to be subject to an independent theoretical error
which is implemented by smearing the mean value for each channel separately
and then combining the smeared channels.
The corresponding probability, for the simple case in which the $\ordQCDtwo$
background is absent,is given by:

\begin{equation}
P_U(\vec{k};\vec{\lambda}) = \prod_i\int dx_i \rho(x_i) \mathcal{P}(k_i,(1+x_i)\lambda_i)
\label{eq:Puncorr}
\end{equation}

where $\mathcal{P}(k,\lambda)$ is the standard Poisson distribution with mean $\lambda$ 
and

\begin{equation}
\rho(x) = \left\{
\begin{array}{rl}
\frac{1}{2\times 0.3} & \mbox{if } |x| < 0.3\\
0 & \mbox{otherwise} \\
\end{array} \right.
\end{equation}

models the (flat) theoretical uncertainty.

Combining separately the two set of channels
we obtain the
probabilities to exclude the SM at 95\%CL and at 99.7\%CL
shown in \tbn{tab:pbsm_nocor}. 

%\begin{table}
%\centering
\TABLE{
\begin{tabular}{|l|c|c|c|c|}
\hline
\multicolumn{5}{|c|}{\bf uncorrelated error} \\
\hline
	& \multicolumn{2}{|c|}{\bf non-reconstructable} 
	                                & \multicolumn{2}{|c|}{\bf reconstructable}\\
\hline
	& NOH		& SILH	 	& NOH	     & SILH \\
\hline
95\%CL  & $>$99.99 \%	& 76.24 \% 	& 99.96\%  & 52.81\% \\
\hline
99.7\%CL& 99.98 \%	& 40.34\%     & 99.37\%  & 18.61\%  \\
\hline
\end{tabular}
\label{tab:pbsm_nocor}
\caption{Probability to exclude the SM with different confidence levels, for different strong 
alternative scenarios combining the three non-reconstructable channels 
($2j\ell^\pm\ell^\pm\nu\nu$, $ZZ\rightarrow 2j\ell\nu\ell\nu$ and $WW\rightarrow 2j\ell\nu\ell\nu$),
and the four reconstructable ones ($4j\ell\nu$, $4j\ell\ell$, $2j3\ell\nu$ and $2j4\ell$).
Theoretical errors are assumed to be uncorrelated as in \eqn{eq:Puncorr}.}
%\end{table}
}

%\begin{table}
%\centering
\TABLE{
\begin{tabular}{|l|c|c|c|c|}
\hline
\multicolumn{5}{|c|}{\bf strongly-correlated error} \\
\hline
	& \multicolumn{2}{|c|}{\bf non-reconstructable} 
	                                & \multicolumn{2}{|c|}{\bf reconstructable}\\
\hline
	& NOH		& SILH	    	& NOH	    & SILH \\
\hline
95\%CL  & 99.99 \%	& 66.05\%	& 99.34\% & 44.07\%  \\
\hline
99.7\%CL& 99.66 \%	& 34.33\%	& 94.24\% & 16.07\%  \\
\hline
\end{tabular}
\caption{Probability to exclude the SM with different confidence levels, for different strong 
alternative scenarios combining the three non-reconstructable channels 
($2j\ell^\pm\ell^\pm\nu\nu$, $ZZ\rightarrow 2j\ell\nu\ell\nu$ and $WW\rightarrow 2j\ell\nu\ell\nu$),
and the four reconstructable ones ($4j\ell\nu$, $4j\ell\ell$, $2j3\ell\nu$ and $2j4\ell$).
Theoretical errors are assumed to be fully correlated as in \eqn{eq:Pstrcorr}.}
\label{tab:pbsm_strcorr}
%\end{table}
}

The hypothesis that the theoretical errors in each channel are independent
may underestimate the actual uncertainty. The dominant production mechanisms
are the same for all channels and therefore it is likely that
both pdf and scale uncertainties are fully correlated between channels.
Assuming this to be the case for the total theoretical uncertainty,
the average values of each channel which enter the combination
must be shifted by a common factor
and the corresponding probability $P_C$ can be expressed as

\begin{equation}
P_C(\vec{k};\vec{\lambda}) = \int dx \rho(x) \prod_i \mathcal{P}(k_i,(1+x)\lambda_i).     
\label{eq:Pstrcorr}
\end{equation}

The PBSM@95\%CL and PBSM@99.7\%CL in case of complete correlation between
theoretical errors are shown in \tbn{tab:pbsm_strcorr}.
Comparing \tbn{tab:pbsm_strcorr} with \tbn{tab:pbsm_nocor}, it is clear that
dropping the hypothesis of independent theoretical errors degrades little
the overall probability. Within the present theoretical framework it remains certain
that the no--Higgs case would be distinguished from the SM case.
In the SILH model the PBSM@95\%CL drops from 76\% to 66\% for the 
non-reconstructable channels and from 53\% to 44\% for the reconstructable ones.

The non--reconstructable channels presented here, 
independently of the detailed treatment of the theoretical uncertainty,
provide a
better discrimination between the SM and the BSM scenarios, despite their
low rates, than those in which the invariant mass of the boson pair
can be measured. This is clearly related to the absence of huge QCD backgrounds,
which are instead present in the $V+4j$ channels. The statistical uncertainties
of these background are large and spoil the significance of the corresponding
channels even when the backgrounds are assumed to be measured from the sidebands
of the weak boson peak in the mass distribution of the two central jets and then
subtracted as proposed in \rfs{Ballestrero:2008gf,Ballestrero:2009vw}.

\begin{figure}[ht!]
\centering
\subfigure{	 
\hspace*{-2.1cm} 
\includegraphics*[width=8.3cm,height=6.2cm]{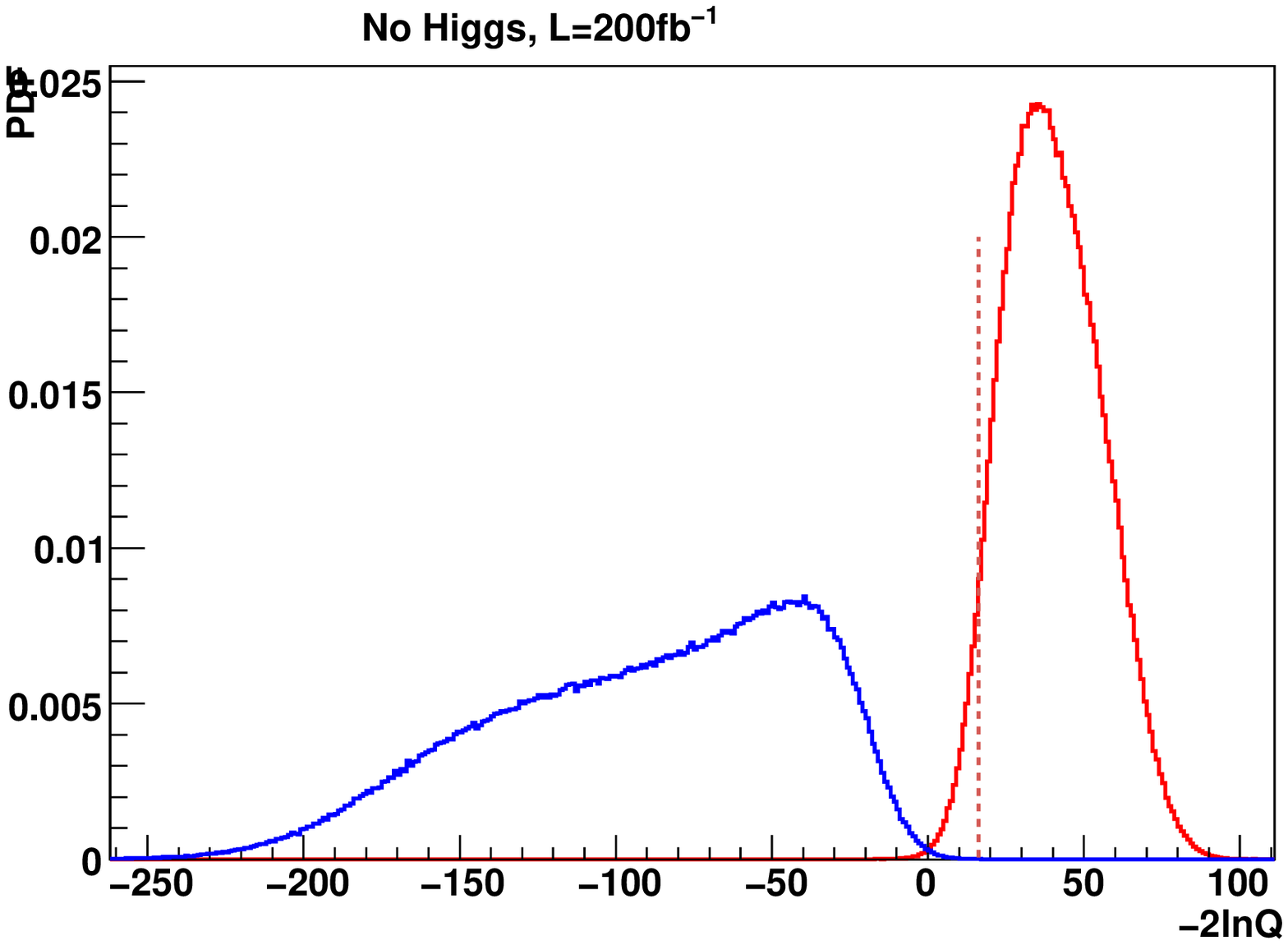}
\hspace*{-0.7cm}
\includegraphics*[width=8.3cm,height=6.2cm]{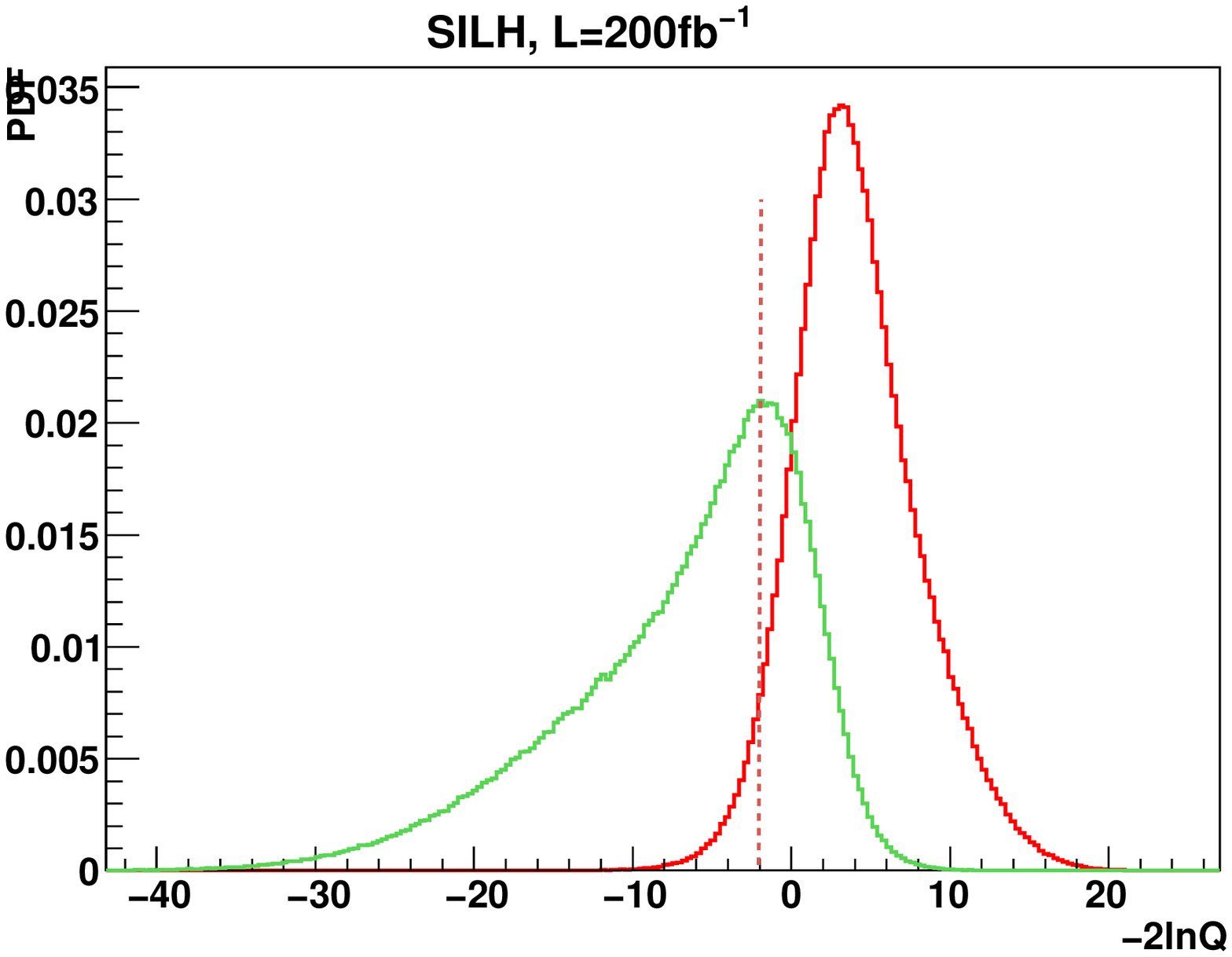}
\hspace*{-3cm}
}
\vspace{-0.4cm}
\subfigure{
\hspace*{-2.1cm}
\includegraphics*[width=8.3cm,height=6.2cm]{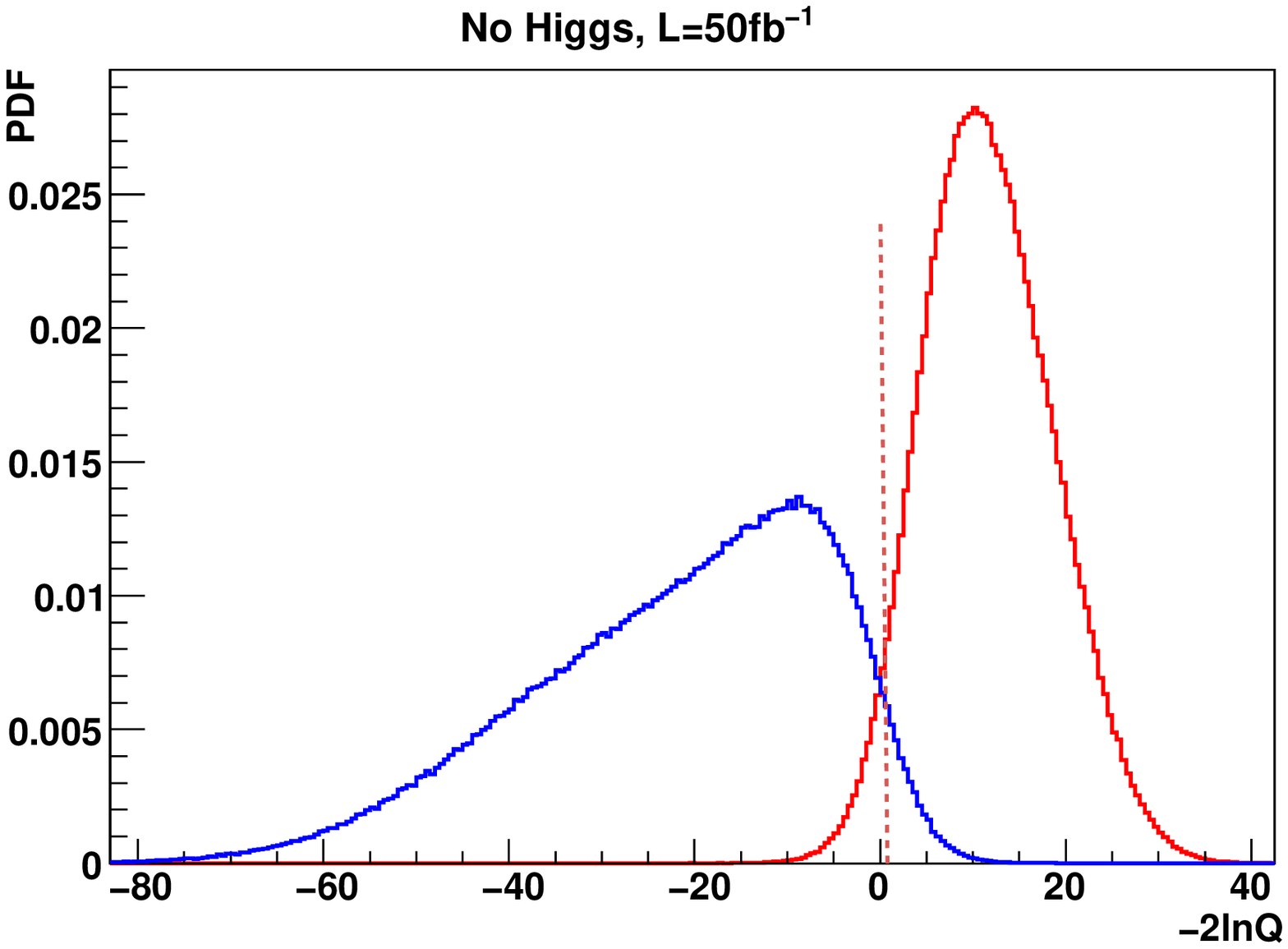}
\hspace*{-0.7cm}
\includegraphics*[width=8.3cm,height=6.2cm]{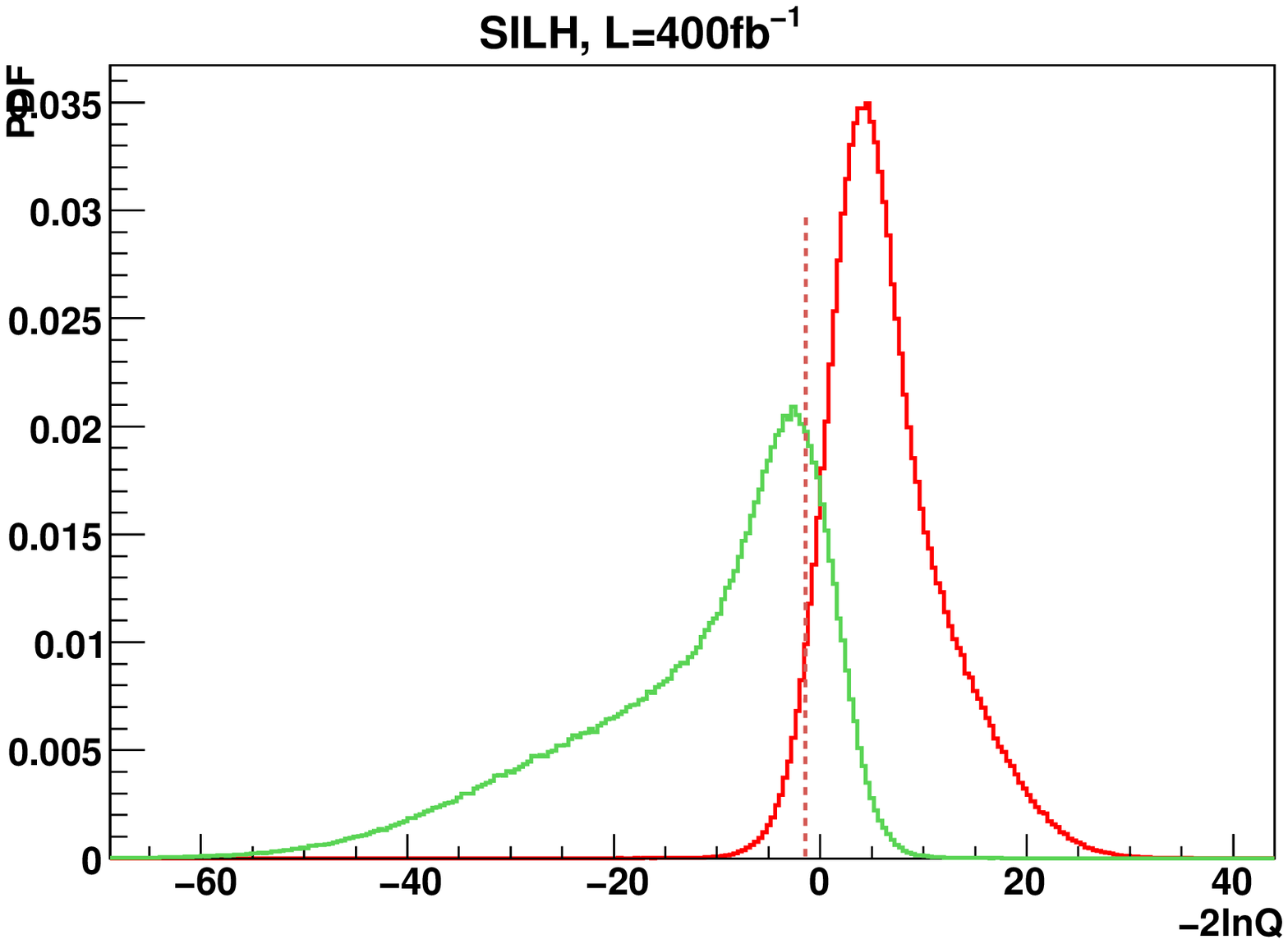}
\hspace*{-3cm}
}
\vspace{0.2cm}
\caption{Combination of all the seven channels using $-2\ln Q$ for the no--Higgs and SILH cases 
using strongly correlated theoretical errors. In the bottom plots 
an integrated luminosity of $L=50\ifb$ is assumed for the no--Higgs scenario and of 
$L=400\ifb$ for the SILH case. For the no--Higgs analysis at $L=50\ifb$
we have used the lowest mass cut for
each channel in order to increase the number of events at this luminosity.
Hence in this case $M_{cut}= 200,\, 300,\, 300,\, 300 \mbox{ GeV}$ for the
$2j\ell^\pm\ell^\pm\nu\nu$, $ZZ\rightarrow 2j\ell\nu\ell\nu$, $WW\rightarrow 2j\ell\nu\ell\nu$
and $2j4\ell$ channel respectively.
For the remaining channels $M_{cut}$ has been kept at $600 \mbox{ GeV}$.
}
\label{fig:lnq_7ch}
\end{figure}

%\begin{table}
%\centering
\TABLE{
\begin{tabular}{|l|c|c|c|c|}
\hline
\multicolumn{5}{|c|}{\bf total combination} \\
\hline
	& NOH		& SILH	 	& NOH($L=50\ifb$) & SILH($L=400\ifb$) \\
\hline
95\%CL  & $>$99.99\%	& 69.32\%	& 96.31\%	  & 80.82\%  	  \\
\hline
99.7\%CL& 99.96\%	& 41.67\% 	& 83.64\%	  & 57.16\%  	  \\
\hline
\end{tabular}
\caption{Statistical combination of all seven channels.
Also shown are the results 
using an integrated luminosity of $L=50\ifb$ for the no--Higgs and 
of $L=400\ifb$ for the SILH scenario. For the no--Higgs analysis we have used the lowest mass cut for
each channel in order to increase the number of events at this luminosity.
}
\label{tab:pbsm_lnq_7ch}
%\end{table}
}

The distribution of $-2 \ln Q$, for the combination of all seven channels,
in the Higgsless and SILH scenarios with our standard
luminosity are shown in the top row of \fig{fig:lnq_7ch}.
In the bottom row of \fig{fig:lnq_7ch} we present the distributions for the Higgsless case
with a luminosity of $50\ifb$ and for the SILH model with $L=400\ifb$.
In all cases theoretical errors are treated as fully correlated.
For the no--Higgs case at $L=50\ifb$ we have adopted the lowest invariant mass cut
reported in Tables
\ref{tab:res_llss_1}, \ref{tab:res_zz_1}, \ref{tab:res_ww_1} and \ref{tab:res_4l_1}
for the reactions discussed in this paper,
in order to increase the number of events at this luminosity.
Explicitly,
$M_{cut}= 200,\, 300,\, 300,\, 300 \mbox{ GeV}$ for the
$2j\ell^\pm\ell^\pm\nu\nu$, $ZZ\rightarrow 2j\ell\nu\ell\nu$, $WW\rightarrow 2j\ell\nu\ell\nu$
and $2j4\ell$ channel respectively.
For the remaining channels $M_{cut}$ has been kept at $600 \mbox{ GeV}$.
The corresponding PBSM@95\%CL and PBSM@99.7\%CL are given in \tbn{tab:pbsm_lnq_7ch}.
Assuming an integrated luminosity of $200\ifb$ the PBSM@95\%CL for the SILH case is about
69\% which increases to 80\% if the luminosity is doubled.
The corresponding figures for the PBSM@99.7\%CL are 42\% and 57\% respectively.
The probability to distinguish at 95\%CL the no--Higgs case from the light Higgs picture
with a reduced luminosity of $L=50\ifb$ remains above 95\%.

%\begin{table}[ht!]
%\centering
\TABLE{
\begin{tabular}{|l|c|c|c|c|c|c|c|c|c|}
\hline
Channel & \multicolumn{3}{|c|}{no Higgs}      & \multicolumn{3}{|c|}{SILH} 	  & SM & $t\bar{t}jj$ & B \\ 
\hline
    & $\sigma$(fb) &  $t\bar{t}jj \subset S$ & $t\bar{t}jj = B$ & $\sigma$(fb) & $t\bar{t}jj \subset S$ & $t\bar{t}jj = B$ & $\sigma$(fb) & $\sigma$(fb) & $\sigma$(fb) \\ 
\hline
$2j\ell \nu \ell \nu$&.0968  & 77.9\% & 85.0\% & .0533 & 22.0\% & 26.6\% & .0332 & .0215 & --\\
\hline
$4j\ell\nu$ & 2.36 & 90.1\% & 96.2\% & 1.41  & 30.9\% & 35.2\% & 1.05 & .463 & 9.78 \\
\hline
\end{tabular}
\caption{
PBSM@95\%CL for $200\ifb$ with $t\bar{t}jj$ as as part of the signal ($t\bar{t}jj \subset S$) and $t\bar{t}jj$
as a background ($t\bar{t}jj = B$).
We also give the total cross section for the $2jW^+W^- \rightarrow 2j\ell^+\ell^-\nu\nu$ 
and $4jl\nu$ channels with the full set of cuts and $M_{cut} = 600  \mbox{ GeV}$.
}
\label{tab:res_ww_ttjets}
%\end{table}
}

%\begin{table}[ht!]
%\centering
\TABLE{
\begin{tabular}{|c|c|c|c|c|c|c|}
\hline
     & \multicolumn{2}{|c|}{non-reconstructable} & \multicolumn{2}{|c|}{reconstructable} & \multicolumn{2}{|c|}{all channels}\\ 
\hline
 scenario        & $t\bar{t}jj = B$ & $t\bar{t}jj \subset S$ & $t\bar{t}jj = B$ & $t\bar{t}jj \subset S$ & $t\bar{t}jj = B$ & $t\bar{t}jj \subset S$\\ 
\hline
SILH             &  64.93\% & 63.25\% & 44.07\% & 40.46\% & 68.38\% & 67.47\% \\
\hline
NOH 		 & 99.98\% & 99.97\% & 99.34\%  & 98.27\% & $>$99.99\%  & $>$99.99\% \\
\hline
\end{tabular}
\caption{
PBSM@95\%CL combining the non-reconstructable channels, the reconstructable ones and finally all channels
with $t\bar{t}jj$ as as part of the signal ($t\bar{t}jj \subset S$) and $t\bar{t}jj$
as a background ($t\bar{t}jj = B$).
For the $2jW^+W^- \rightarrow 2j\ell^+\ell^-\nu\nu$ process  $M_{cut} = 500  \mbox{ GeV}$ has been used.
}
\label{tab:comb_nonrec}
%\end{table}
}

Before stating our conclusions we return for completeness to the possible modification to our results
if the $t\bar{t}jj$ turns out not to be measurable, even though as discussed previously we do not regard
this prospect as probable. In this case the contribution from $t\bar{t}jj$ production should be considered
as part of the signal S and therefore subject to theoretical uncertainties in addition to the
statistical ones. For simplicity we have assumed the same range of variation for this
process as for all others involved. We refer to this possibility as $t\bar{t}jj$ as signal ($t\bar{t}jj \subset S$)
while the framework in which $t\bar{t}jj$ production is considered as a measured and extrapolated background
is described as $t\bar{t}jj$ as background ($t\bar{t}jj = B$).
In \tbn{tab:res_ww_ttjets} we present the cross sections and the PBSM@95\%CL for the two channels
which are affected by the $t\bar{t}jj$ background, namely $2jW^+W^-$ and $4jW$.
In \tbn{tab:comb_nonrec} we compare the two results for the different combinations of channels. 
While there is a noticeable decrease of the PBSM for the individual reactions in \tbn{tab:res_ww_ttjets},
the overall combinations are hardly affected.

These results suggest that the no--Higgs scenario can be disproved with a rather modest 
luminosity. This implies that any model which predicts vector vector scattering rates
larger than those in the no--Higgs case can be disproved or verified with the
same luminosity.

Our conclusions for the SILH framework are less optimistic. Clearly a substantial increase
in luminosity and a combination of the results obtained by ATLAS and CMS are highly desirable.
Furthermore, it should be recalled that the particular instance of SILH model
we have discussed is a rather extreme case and that for smaller values of $c_H\xi$
results even closer to the SM ones are expected.

%%%%%%%%%%%%%%%%%%%%%%%%%%%%%%%%%%%%%%%%%%%%%%%%%%%%%%%%%%%%%%%%%%%%%%%%%
\section{Conclusions}
\label{sec:conclusions}

We have examined in detail at parton level the processes 
$2\ell 2\nu 2j$ and $4\ell 2j$, $\ell = \mu,\, e$ including all
irreducible backgrounds contributing to these six parton final states. We have
considered three scenarios: a light Higgs SM framework with $M_H = 200 \mbox{
GeV}$, one instance of the SILH models and an infinite mass Higgs scenario in
order to determine whether the two BSM models can be distinguished from the SM
at the LHC using boson--boson scattering.
Because of the absence of large QCD backgrounds,
the non--reconstructable channels presented here provide a
better discrimination between the SM and the BSM scenarios, despite their
low rates, than those in which the invariant mass of the boson pair
can be measured.

The results for the channels discussed above
have been combined with those obtained in
\rf{Ballestrero:2008gf} for $4j\ell\nu$ production
and those obtained in \rf{Ballestrero:2009vw} for $4j\ell^+\ell^-$ and $2j3\ell\nu$.
We have estimated the total probability, in
the two BSM scenarios, of finding a result
outside the 95\% probability range in the Standard Model. This probability turns
out to be essentially 100\% for the Higgsless case and 69\% for the SILH model.
These probabilities correspond to an integrated luminosity of $L = 200 \mbox{
fb}^{-1}$ and to the sum of all electron and muon channels.

\section *{Acknowledgments}

A.B. wishes to thank the Dep. of Theoretical Physics of Torino University
for support. \\
This work has been supported by MIUR under contract 2006020509 004 and by the
European Community's Marie-Curie Research Training Network under contract
MRTN--CT--2006--035505 Tools and Precision Calculations for Physics Discoveries
at CollidersÕ

%\hfill*
%\eject
%\newpage
\vspace{2cm}

%%%%%%%%%%%%%%%%%%%%%%%%%%%%%%%%%%%%%%%%%%%%%%%%%%%%%%%%%%%%%%%%%%%%%%%%%

\end{document}